\documentclass[twocolumn]{aastex631}
\usepackage{amssymb}
\usepackage{amsmath}
\usepackage{comment}
\usepackage[colorlinks=true]{hyperref}
\usepackage{cleveref}
\usepackage{mathtools}

\shorttitle{Principle of Inhomogeneity III}
\shortauthors{Zhang}
\graphicspath{{./}{figures/}}

\begin{document}

\title{The Inhomogeneity Effect III: \\
Weather Impacts on the Heat Flow of Hot Jupiters}

\correspondingauthor{Xi Zhang}
\email{xiz@ucsc.edu}

\author{Xi Zhang}
\affiliation{Department of Earth and Planetary Sciences, University of California Santa Cruz, Santa Cruz, CA 95064, USA}
\author{Cheng Li}
\affiliation{Department of Climate and Space Sciences and Engineering, University of Michigan, Ann Arbor, MI 48109 USA}
\author{Huazhi Ge}
\affiliation{Department of Earth and Planetary Sciences, University of California Santa Cruz, Santa Cruz, CA 95064, USA}
\affiliation{51 Pegasi b Fellow}
\author{Tianhao Le}
\affiliation{Department of Earth and Planetary Sciences, University of California Santa Cruz, Santa Cruz, CA 95064, USA}

\begin{abstract}

The interior flux of a giant planet impacts atmospheric motion, and the atmosphere dictates the interior's cooling. Here we use a non-hydrostatic general circulation model (Simulating Nonhydrostatic Atmospheres on Planets, SNAP) coupled with a multi-stream multi-scattering radiative module (High-performance Atmospheric Radiation Package, HARP) to simulate the weather impacts on the heat flow of hot Jupiters. We found that the vertical heat flux is primarily transported by convection in the lower atmosphere and regulated by dynamics and radiation in the overlying ``radiation-circulation" zone. The temperature inversion occurs on the dayside and reduces the upward radiative flux. The atmospheric dynamics relay the vertical heat transport until the radiation becomes efficient in the upper atmosphere. The cooling flux increases with atmospheric drag due to increased day-night contrast and spatial inhomogeneity. The temperature dependence of the infrared opacity greatly amplifies the opacity inhomogeneity. Although atmospheric circulation could transport heat downward in a narrow region above the radiative-convective boundary, the opacity inhomogeneity effect overcomes the dynamical effect and leads to a larger overall interior cooling than the local simulations with the same interior entropy and stellar flux. The enhancement depends critically on the equilibrium temperature, drag, and atmospheric opacity. In a strong-drag atmosphere hotter than 1600 K, a significant inhomogeneity effect in three-dimensional (3D) models can boost interior cooling several-fold compared to the 1D radiative-convective equilibrium models. This study confirms the analytical argument of the inhomogeneity effect in \citet{zhangInhomogeneityEffectInhomogeneous2023,zhangInhomogeneityEffectII2023}. It highlights the importance of using 3D atmospheric models in understanding the inflation mechanisms of hot Jupiters and giant planet evolution in general.

\end{abstract}



\section{Introduction} \label{introsec}

Traditional evolution models cannot explain the large sizes of some hot Jupiters with high equilibrium temperatures ($T_{eq}>1000~\mathrm{K}$, \citealt{demoryLackInflatedRadii2011,laughlinAnomalousRadiiTransiting2011,millerHeavyelementMassesExtrasolar2011}). The bloated size of these planets suggests a higher entropy interior than what should be expected for their age. Furthermore, observations reveal that the radius excess correlates more closely with incident stellar flux than orbital distance or period \citep{laughlinAnomalousRadiiTransiting2011,weissMassKOI94dRelation2013,thorngrenBayesianAnalysisHotJupiter2018,thorngrenSlowCoolingFast2021}. This correlation casts doubt on the efficacy of the inflation mechanisms at work in the interior, such as convective inhibition (e.g., \citealt{chabrierHeatTransportGiant2007,leconteNewVisionGiant2012}) and tidal heating (e.g., \citealt{bodenheimerTidalInflationShortPeriod2001,bodenheimerRadiiExtrasolarGiant2003,guEffectTidalInflation2003,winnObliquityTidesHot2005,jacksonTidalHeatingExtrasolar2008,liuConstraintsDeepseatedZonal2008,ibguiCoupledEvolutionTides2009,millerInflatingDeflatingHot2009,leconteTidalHeatingSufficient2010,ibguiTidalHeatingModels2010,ibguiExplorationsViabilityCoupled2011,kurokawaRadiusAnomalyHot2015}). For example, it appears unlikely that eccentricity tides can explain the distributions of inflated hot Jupiters \citep{laughlinMassRadiusRelationsGiant2018,fortneyHotJupitersOrigins2021}. Instead, atmospheric processes interacting with stellar irradiation may be the key to solving this size mystery.

\subsection{Importance of Atmospheric Inhomogeneity on Cooling Hot Jupiters} \label{intro}
As the outer boundary of a giant planet, the atmosphere regulates incoming and outgoing heat flow. A higher level of stellar irradiation would cause the radiative-convective boundary (RCB) to move deeper into the atmosphere and reduce the cooling flux from the interior (e.g., \citealt{guillotGiantPlanetsSmall1996}). However, this effect alone is insufficient to explain the observed size anomalies. Modifying the atmospheric opacity can further reduce cooling (\citealt{burrowsPossibleSolutionsRadius2007}), but most proposed solutions involve redistributing energy from the upper atmosphere to the deeper interior, which can be broadly defined as the adiabatic region below the lowest RCB. These mechanisms include Ohmic heating (\citealt{batyginInflatingHotJupiters2010,pernaOhmicDissipationAtmospheres2010,pernaEffectsIrradiationHot2012,batyginEvolutionOhmicallyHeated2011,huangOhmicDissipationInteriors2012,rauscherThreedimensionalAtmosphericCirculation2013,wuOhmicHeatingSuspends2013,rogersMagneticEffectsHot2014}), atmospheric thermal tides on asynchronous hot Jupiters \citep{arrasThermalTidesFluid2010,guModelingThermalBulge2019}, turbulent kinetic energy transport (\citealt{showmanAtmosphericCirculationTides2002, guillotEvolution51Pegasus2002, youdinMechanicalGreenhouseBurial2010}), and downward heat advection by atmospheric circulation and waves (\citealt{showmanAtmosphericCirculationTides2002, guillotEvolution51Pegasus2002, tremblinAdvectionPotentialTemperature2017,mendonca2020,sainsbury-martinezIdealisedSimulationsDeep2019,sainsbury-martinezEvidenceRadiusInflation2023}).

The deposited heating efficiency---the ratio of deposited energy to stellar irradiation---can provide useful constraints on the mechanisms. To explain the observed radius anomalies, the heating efficiency is typically a few percent, depending on the depth and duration of energy deposition (e.g., \citealt{guillotEvolution51Pegasus2002,spiegelThermalProcessesGoverning2013,ginzburgHotJupiterInflationDue2015,komacekStructureEvolutionInternally2017,thorngrenBayesianAnalysisHotJupiter2018,thorngrenIntrinsicTemperatureRadiativeConvective2019}). Assuming that the heat is deposited deep and that the current state of hot Jupiters is in energy equilibrium, this allow us to connect the hot interior entropy to the atmosphere and calculate the net cooling flux at the top of the atmosphere (TOA) to infer the required heating efficiency of inflated hot Jupiters. \citet{thorngrenBayesianAnalysisHotJupiter2018} combined hot Jupiter data and a 1D structure model to estimate heating efficiency as a function of the equilibrium temperature ($T_{eq}$). They suggested that the heating efficiency peaks at around $T_{eq}\sim1600~\mathrm{K}$ and decreases towards cooler and hotter planets. This finding appears to be consistent with predictions from the Ohmic heating mechanism. A later analysis in \citet{sarkisEvidenceThreeMechanisms2021} found that the heating efficiency reaches a maximum of 2.5\% at about 1860 K. It seems that Ohmic dissipation, advection of potential temperature, and thermal tides are all consistent with their efficiency distribution. 

Both above studies adopted the 1D radiative-convective-equilibrium (RCE) atmospheric models to estimate the planetary cooling flux assuming full heat redistribution between the dayside and nightside. However, a 3D atmosphere is usually not in RCE and is not homogeneous. Some additional effects are essential for estimating the cooling rate of the planet. First, the inhomogeneity of incoming stellar radiation from the 3D spherical geometry is significant on hot Jupiters that undergo extreme irradiation patterns with large day-night and equator-pole temperature differences. It has been recognized that the day-night and equator-pole contrast can affect the evolution of hot Jupiters (e.g., \citealt{guillotEvolution51Pegasus2002, budajDayNightSide2012,spiegelThermalProcessesGoverning2013,rauscherINFLUENCEDIFFERENTIALIRRADIATION2014}). Second, because atmospheric opacity depends on temperature, a non-uniformly distributed opacity would also change the cooling rate. 

In \citet{zhangInhomogeneityEffectInhomogeneous2023}, we proposed a general principle that the inhomogeneity in planetary surfaces and atmospheres would generally accelerate planetary cooling in various types of planets. The average internal heat fluxes between two inhomogeneous columns can be much larger than the uniform column with the average stellar irradiation or opacity for giant planets. The difference can reach a factor of a few for the irradiation inhomogeneity and more than an order of magnitude for the opacity inhomogeneity. In \citet{zhangInhomogeneityEffectII2023}, we further generalized the theory and found that the inhomogeneity caused by orbital and rotational configurations can significantly affect the internal heat flux of giant planets.

In principle, the inhomogeneity of an atmosphere is determined by many processes including radiation and dynamics. Previous models primarily focus on radiation. \citet{zhangInhomogeneityEffectInhomogeneous2023,zhangInhomogeneityEffectII2023} explored the inhomogeneity effect using a local RCE model with a simplified treatment of the dynamics. Whereas the general circulation and wave mixing redistribute energy and significantly regulate heating and cooling in the atmosphere. For example, fast jets could smooth out zonal inhomogeneity, but waves could perturb the temperature pattern. The spatial contrast between the dayside and nightside could be large in the presence of strong drag—which might be relevant to the Ohmic heating mechanism (e.g., \citealt{rauscherThreedimensionalAtmosphericCirculation2013,rogersMagneticEffectsHot2014}). Furthermore, atmospheric dynamics can transport energy downward via circulation and wave mixing, leading to interior heating  (e.g., \citealt{showmanAtmosphericCirculationTides2002,youdinMechanicalGreenhouseBurial2010, tremblinAdvectionPotentialTemperature2017,mendonca2020}). To account for the rich behavior and complex interactions in a dynamical atmosphere, a 3D atmospheric simulation is needed to self-consistently explore weather effects on the heat flow of hot Jupiters, which are otherwise overlooked in 1D models.

3D general circulation models (GCM) have proven to be successful tools in interpreting exoplanet data over the past decades (see reviews in \citealt{hengAtmosphericDynamicsHot2015,showmanAtmosphericDynamicsHot2020,zhangAtmosphericRegimesTrends2020,fortneyHotJupitersOrigins2021}). However, only a few studies are dedicated to studying heat flow in connection with a hot interior in the context of heating or cooling mechanisms. \citet{showmanAtmosphericCirculationTides2002} pioneered the analysis of energy transport on hot Jupiters utilizing a GCM. They proposed that kinetic, potential, and thermal energy could be conveyed downward through atmospheric dynamics, akin to the Walker circulation observed on Earth. \citet{mendonca2020} analyzed the angular momentum and heat transport of hot Jupiters using a non-hydrostatic GCM and found that the heat can be transported downward by mean circulation and stationary waves. \citet{hammondRotationalDivergentComponents2021} and \citet{lewisTemperatureStructuresAssociated2022} analyzed the impact of the horizontal flow pattern on the heat transport on a canonical hot Jupiter. \cite{komacekEffectInteriorHeat2022} showed that a high-entropy interior and a large heat flux produce different horizontal temperature and flow structures than a low-entropy case. Simulations including the magnetic effect in the Ohmic heating mechanism, either with simple drag schemes \citep{rauscherRoleDragEnergetics2012,rauscherThreedimensionalAtmosphericCirculation2013,rauscherAtmosphericCirculationObservable2014,beltzExploringEffectsActive2022} or using complex MHD models \citep{rogersMagnetohydrodynamicSimulationsAtmosphere2014,rogersMagneticEffectsHot2014,rogersConstraintsMagneticField2017}, show significantly different temperature patterns with large spatial inhomogeneity compared with traditional drag-free hydrodynamic simulations (e.g., \citealt{showmanAtmosphericCirculationHot2009}). \cite{sainsbury-martinezIdealisedSimulationsDeep2019} used a GCM with the Newtonian cooling scheme to evaluate inflation mechanisms and showed that the interior could be heated up by atmospheric circulation. Using another GCM with a radiative transfer scheme from \cite{schneiderNoEvidenceRadius2022}, recent work in \cite{sainsbury-martinezEvidenceRadiusInflation2023} confirmed that heat can be transported downward by large-scale circulations on hot Jupiter WASP-76b. However, their total energy fluxes appeared to exhibit large vertical variations (see their Figure 3). It would be important to analyze the vertical transport of energy fluxes in hot Jupiter atmospheres with a more energy-conserving scheme.

\subsection{Methodology of This Study}

The aim of this study is to shed light on the internal heat flux transport through the weather layer on hot Jupiters by employing atmospheric dynamics models. We chose a non-hydrostatic (fully compressible) GCM featuring grey radiative transfer to enable the resolution of convection. By assuming that the temperature is horizontally homogenized over isobars (constant pressure levels) in the deep interior due to strong convection, we connected the weather layer model to the deep interior, fixing the temperature at the bottom of the models to a constant value. This approach facilitated a comparison of net cooling fluxes at the TOA for local 2D/3D models with minimal spatial inhomogeneity, and a global 3D model in a tidally locked configuration, all having the same total incoming stellar flux and the same bottom temperature.

We followed several technical steps in this work:

First, for each equilibrium temperature, we drew upon the heating efficiency required to explain the inflated hot Jupiter size based on previous 1D models (e.g., \citealt{thorngrenBayesianAnalysisHotJupiter2018,sarkisEvidenceThreeMechanisms2021}). This heating efficiency, when multiplied with the incoming stellar flux, yielded the additional heating flux necessary to inflate the planet. Assuming a steady state for the planet (meaning no further evolution of the planetary size), the internal flux at the TOA must be equivalent to this additional heat flux.

We then implemented a local 2D/3D radiative-dynamical model with a fixed bottom temperature. We chose a bottom pressure of 200 bars, deep within the convective zone, such that the bottom temperature defines the entropy of the entire planetary interior. We adjusted the deep temperature to match the net cooling flux at the TOA (i.e., internal heat flux) to the heating flux derived from the 1D models in \citet{thorngrenBayesianAnalysisHotJupiter2018} and \citet{sarkisEvidenceThreeMechanisms2021}. Since our grey opacity differs from their non-grey opacity, we had to adjust the bottom temperature in our local model to generate the same outgoing TOA fluxes. The internal heat flux at the TOA is partially influenced by the energy sink/source related to the temperature relaxation and wind drag at the deep boundary.

We applied the same deep temperature from the local model to the global 3D GCM and simulated the weather and heat transport in a tidally locked configuration. By analyzing the internal heat flux at the TOA in this 3D global model and comparing it with the local model flux, we sought to better understand the impact of radiative and dynamical processes in the 3D atmosphere on the internal heat transport of hot Jupiters. We specifically discussed whether the downward dynamical transport of heat flux could provide a satisfactory explanation for size inflation. If so, the internal heat flux at the TOA would be zero in our model.

Lastly, we applied the atmospheric drag in the global simulations to change the spatial contrast. We altered the drag timescale to study the inhomogeneity effect induced by drag. By considering a range of equilibrium temperatures, we analyzed how the inhomogeneity effect fluctuates with planetary temperature. In addition, we incorporated both temperature-independent and temperature-dependent opacity in the infrared. This helped us comprehend the effect of temperature inhomogeneity on infrared opacity and cooling flux, leading to a re-evaluation of the required heating efficiency of inflated hot Jupiters.
 
The paper is structured as follows. In Section \ref{sec:model}, we describe the local and global models and the opacity sources. Then we first show the results and fluxes in the local models with $T_{eq}=1600$ K in Section \ref{sec:local}, followed by a long Section \ref{sec:global} to highlight the important features in the weather pattern, heat flux, and RCB in the 3D global simulations with and without drag. To further diagnose the detailed mechanisms and highlight the opacity inhomogeneity effect, we perform simulations with different opacities in Section \ref{sec:opa}. Finally, we discuss the implications of this study in the context of observations and interior heating mechanisms for hot Jupiters in Section \ref{sec:implication}. We investigate the system behavior with different drag timescales and equilibrium temperatures. We summarize the main points in Section \ref{sec:sum}.

\section{Model Description}\label{sec:model}

All local and global simulations in this study were carried out using the same atmospheric model, which has two main components: a dynamic core SNAP (Simulating Nonhydrostatic Atmospheres on Planets, \citealt{liSimulatingNonhydrostaticAtmospheres2019, geGlobalNonHydrostaticAtmospheric2020}) and a multi-stream multiple-scattering radiative transfer solver HARP (High-performance Atmospheric Radiation Package, \citealt{liHighperformanceAtmosphericRadiation2018}). The model is built on top of the Athena++ framework (\citealt{whiteConsistentApproximateModels2005,stoneAthenaAdaptiveMesh2020}), allowing it to exploit the features provided by the Athena++ code, including static/adaptive mesh-refinement, curvilinear geometry, and dynamic task scheduling. Athena++ has an excellent ability to scale in parallel computation, a crucial characteristic for computationally expensive simulations. The 3D simulations can be efficiently paralleled to thousands of CPUs on supercomputers. Our GCM can also simulate cloud formation with active tracer transport and moist convection (\citealt{geStableSuperadiabaticWeather2023}). We focus on the clear-sky simulations of hot Jupiters in this study.

\subsection{Dynamical Core: SNAP}
SNAP is a non-hydrostatic dynamical solver, designed explicitly for atmospheric simulations. It can simulate moist convection, strong updrafts, and tracer transport from the first principles. SNAP includes a Low Mach number Approximate Riemann Solver (LMARS, \citealt{chenControlVolumeModelCompressible2013}), a fifth-order weighted essentially non-oscillatory (WENO) reconstruction method for subgrid reconstruction (\citealt{shuHighorderFiniteDifference2003}), and a third-order total variation diminishing Runge-Kutta method for time-stepping (see details in \citealt{liSimulatingNonhydrostaticAtmospheres2019,geGlobalNonHydrostaticAtmospheric2020}). 

Typically, a 3D non-hydrostatic simulation runs slowly because the numerical time step is limited by fast propagating sound waves in the vertical dimension. To overcome this issue, a new horizontal-explicit and vertical-implicit (HEVI) scheme for SNAP has been published (\citealt{geGlobalNonHydrostaticAtmospheric2020}). This scheme does not require traditional numerical stabilizers such as hyperviscosity and divergence damping, which simplifies the implementation and improves numerical stability. The timestep with the implicit scheme is not limited by vertically propagating sound waves, significantly improving computational efficiency by more than a factor of 100, comparable to that of explicit non-hydrostatic models. We have performed rigorous numerical benchmark tests such as the challenging Robert rising bubble test (\citealt{robertBubbleConvectionExperiments1993}), Straka sinking bubble test (\citealt{strakaNumericalSolutionsNonlinear1993}), and gravity wave test (\citealt{skamarockEfficiencyAccuracyKlempWilhelmson1994}). We have also validated the 3D global model against the standard \citet{heldProposalIntercomparisonDynamical1994} test for Earth-like simulations and typical exoplanet simulations (\citealt{geGlobalNonHydrostaticAtmospheric2020}). 

Our 2D and 3D local simulations adopt Cartesian coordinates, while the 3D global simulations use spherical-polar coordinates with polar wedge boundary conditions at poles. To further improve computational efficiency, we have generalized the HEVI scheme to include a horizontal implicit scheme  (Li et al. in prep). The computational efficiency is further boosted with our full implicit (FI) scheme with a time step reaching 10 to 20 folds of the Courant-Friedrichs-Lewy (CFL) criteria. The hot Jupiter simulations with the new FI scheme have been validated against those with the previously published HEVI scheme. 

SNAP has two main advantages that make it particularly suitable for analyzing energy flows in planetary atmospheres. First, as a non-hydrostatic solver, it solves the non-hydrostatic pressure as a prognostic variable, which directly resolves convection. Our model does not need convective adjustment schemes adopted from traditional Earth atmospheric studies and is widely used in hydrostatic models for hydrogen-dominated atmospheres. Since we study the interaction between the radiative and convection zones in a hot Jupiter with high interior entropy, a non-hydrostatic solver is essential for realistically simulating convective behaviors in the atmosphere. Second, SNAP directly solves the energy equation with a finite-volume scheme, unlike most GCMs that solve the potential temperature, which is an entropy-like quantity. This allows us to avoid possible energy conservation issues in other numerical schemes and easily analyze the heat flows in hot Jupiter atmospheres.

\subsection{Radiative Transfer: HARP}

\begin{figure*}
\centering \includegraphics[width=0.85\textwidth]{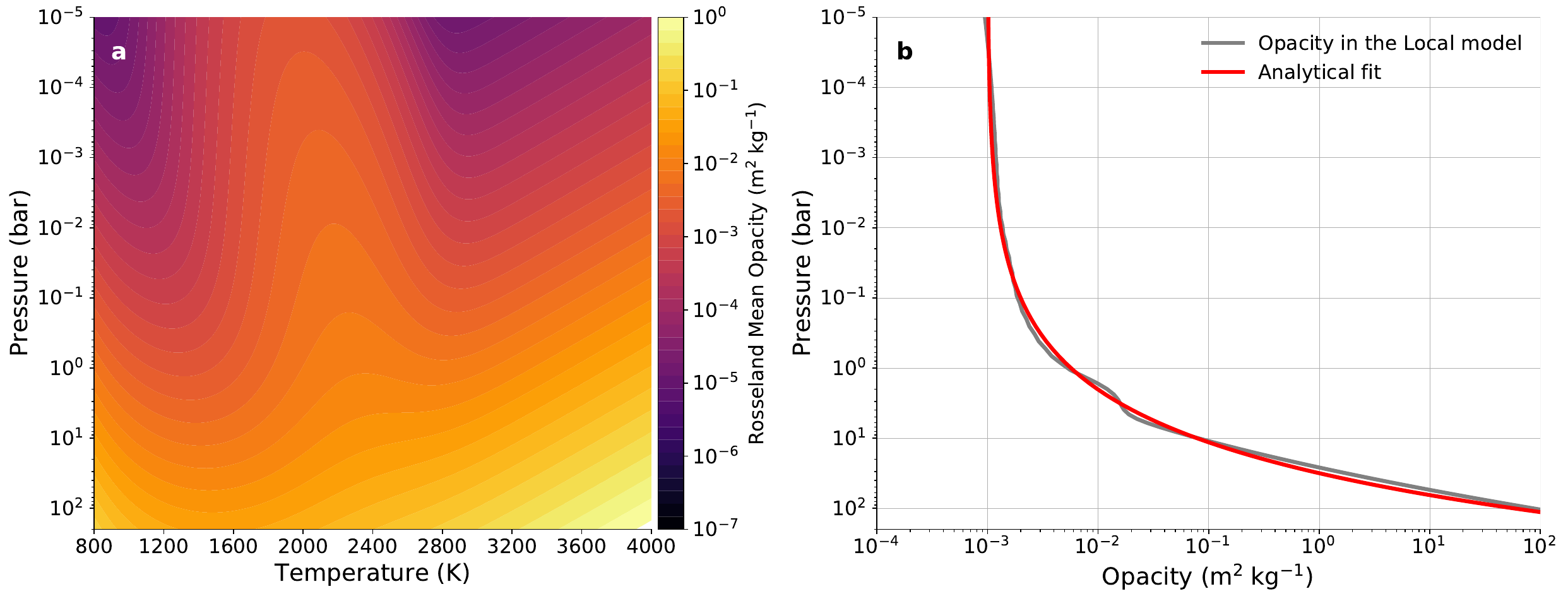}
\caption{High-temperature grey opacity distribution in the simulations. (a) Rosseland-mean opacity as a function of temperature and pressure from \citet{freedmanGaseousMeanOpacities2014}. The metallicity [M/H]=0.5. (b) The domain-averaged Rosseland-mean opacity profile (grey) from the 2D local model with $T_{eq}=1600 K$ and the fitted analytical profile as a function of pressure (red). This analytical opacity profile is used as temperature-independent IR opacity in some simulations in this study.}\label{opa}
\end{figure*}

The radiative transfer module used in our 3D GCM is the High-performance Atmospheric Radiation Package (HARP, \citealt{liHighperformanceAtmosphericRadiation2018}). HARP is based on a multi-stream multi-scattering code DISORT (\citealt{stamnesNumericallyStableAlgorithm1988,burasNewSecondaryscatteringCorrection2011}) to calculate shortwave stellar heating and longwave thermal cooling of the atmosphere, assuming a plane-parallel geometry. One of the highlights of HARP is the multi-stream radiative transfer scheme that allows for a more accurate calculation of the angular flux distribution of the aerosol scattering than the current two-stream scheme in existing GCMs. Using a \textit{correlated-k} approach, HARP has been validated against the line-by-line model in \citet{zhangRadiativeForcingStratosphere2013} for Jupiter and has been applied to other giant planets to update their thermal cooling rates and relaxation timescales (\citealt{liHighperformanceAtmosphericRadiation2018}).

HARP is built on the same infrastructure as Athena++ and is easily coupled with SNAP. However, like all current radiative modules for planetary GCMs, the HARP radiative transfer solver assumes a plane-parallel atmosphere. This assumption may lead to inaccuracies in energy calculations, particularly for spherical, extended atmospheres on inflated hot Jupiters. The simulated domain usually spans vertically about 5 to 10 percent of the planetary radius for a typical hot Jupiter. Under the plane-parallel assumption, a 1D atmospheric column is assumed with an equal area from top to bottom, but in an extended atmosphere, the radiative flux transfers across the columns. Thus, regular plane-parallel transfer solvers, including DISORT, cannot guarantee energy conservation in an extended atmosphere. This is critical for calculating heating and cooling rates, temperature, and outgoing emission flux. 

While solving the radiative transfer equation in a spherical, extended atmosphere is complex and time-consuming (e.g., \citealt{chandrasekharRadiativeEqilibriumStellar1945, mihalasFoundationsRadiationHydrodynamics1999}), we introduced a method to correct the plane-parallel solver to approximate the spherical solver. The details are outlined in Appendix \ref{app:rte}, in which we compared the modified equation to the equation under spherical symmetry. We scaled intensity and thermal radiation source function by the radius square (Equation \ref{tfeqn}). With that, the plane parallel solver calculates the scaled incoming and outgoing fluxes for each atmospheric level. Although our approach could only mimic an asymptotic limit of the spherical solution, it can ensure energy conservation when the radiative flux is transported in a wedge of constant solid angle in the atmosphere. Unlike many GCMs that calculate the net heating rate as a source in the energy equation, the scaled fluxes from HARP are directly used in the finite-volume solver SNAP to conserve energy flux.

We focus on the physical mechanisms of heat flow transport by atmospheric dynamics and radiation in this theoretical study of hot Jupiters. For simplicity, we assumed a cloud-free atmosphere without scattering, for which a two or four-stream scheme in HARP is sufficient. We also assumed a double-grey opacity scheme in which the shortwave stellar heating and longwave thermal cooling are separated into two bands.

According to \citet{zhangInhomogeneityEffectInhomogeneous2023}, the inhomogeneity effect of the thermal IR (longwave) opacity is much stronger than the visible (shortwave) opacity. The IR opacity is a strong function of local atmospheric temperature (\citealt{freedmanGaseousMeanOpacities2014}) that could induce important inhomogeneity effects on the cooling of hot Jupiters. To investigate the impact of the temperature dependence on opacity, we consider two sets of IR grey opacity in our radiative scheme. The first set is a more ``realistic" Rosseland-mean opacity that varies with pressure and temperature. We use the analytical expressions for the gas IR opacity $\kappa_{\rm{IR}}$ (in cm$^2$ g$^{-1}$) provided by \citet{freedmanGaseousMeanOpacities2014}. The opacity is a sum of two components:
\begin{equation}
\kappa_{\rm{IR}} = \kappa_{\text{low-p}} + \kappa_{\text{high-p}}, \label{Topa}
\end{equation}
where
\begin{eqnarray}
\text{log}_{10}\kappa_{\text{low-p}} = c_1\text{tan}^{-1}\left(\text{log}_{10}T-c_2\right) - \\ \frac{c_3}{\text{log}_{10}p+c_4}e^{\left( \text{log}_{10}T-c_5 \right)^2} +c_6\text{met} + c_7 \nonumber
\end{eqnarray}
and
\begin{eqnarray}
\text{log}_{10}\kappa_{\text{high-p}} = c_8 + c_9\text{log}_{10}T + \\ c_{10}(\text{log}_{10}T)^2 + \text{log}_{10}p(c_{11}+c_{12}\text{log}_{10}T) + \nonumber \\ c_{13}\text{met}\left[\frac{1}{2} + \frac{1}{\pi} \text{tan}^{-1} \left( \frac{\text{log}_{10}T - 2.5}{0.2} \nonumber \right) \right].
\end{eqnarray}
$T$ is the temperature in K, $p$ is pressure in dyne cm$^{-2}$, and ``met'' is the metallicity [M/H], which is taken as 0.5 in this study. The coefficients $c_7$ through $c_{13}$ are different in the low-temperature regime ($T<800$ K) and the high-temperature regime ($T>800$ K) and are given in Table 2 in \citet{freedmanGaseousMeanOpacities2014}. \Cref{opa} shows the opacity as a function of temperature and pressure in the high-temperature regime more relevant to hot Jupiters. The opacity generally increases with pressure but shows a large nonlinearity with a temperature when pressure is lower than 1 bar. At the same pressure level (such as 1 mbar), the opacity increases by several orders of magnitude from 1000 K to about 2000 K and then decreases by several orders of magnitude towards 3000 K. The temperature where opacity peaks shifted from 2000 K at high pressure to about 1800 K at lower pressure. Because the temperature variations on tidally locked gas giants are large, this temperature dependence could cause a dramatic opacity inhomogeneity and affect the planet's cooling according to \citet{zhangInhomogeneityEffectInhomogeneous2023}.

The notable rise in opacity around 2000 K is primarily attributed to the presence of TiO/VO gases. However, current observational data does not conclusively confirm the existence of TiO/VO in the atmospheres of planets at these temperatures (e.g., \citealt{mikal-evansDiurnalVariationsStratosphere2022,pelletierVanadiumOxideSharp2023}). From a theoretical standpoint, the TiO/VO gases could be coldly trapped deep within the atmosphere, which could drastically alter their concentration in the photosphere (\citealt{spiegelCanTiOExplain2009,parmentier3DMixingHot2013}). In scenarios where TiO/VO is absent, the relationship between opacity and temperature retains a similar pattern, but the variation around 2000 K is significantly less pronounced. Moreover, at temperatures less than 1800 K, high-temperature clouds would form (\citealt{powellFormationSilicateTitanium2018,gaoAerosolCompositionHot2020}). The opacity of these clouds would fill the minimum of the gas opacity in the 1200-1800K range and naturally moderate the temperature variation of the atmospheric opacity.

Given these uncertainties, we further explore an extreme scenario with an IR opacity independent of temperature in this work. We first calculated the opacity profile based on the domain-averaged temperature profile from our local simulation at $T_{eq}=1600$ K (in Section \ref{sec:local}) and \Cref{Topa}. Then we did a simple analytical fit of the $\kappa$ (in cm$^2$ g$^{-1}$) as a function of $p$ (in dyne cm$^{-2}$), which yields:
\begin{equation}
\log_{10}\kappa_{\rm{IR}} = 0.003 p^{0.4} - 2. \label{TIopa}
\end{equation}

\Cref{opa}b shows the temperature-independent IR opacity profile that generally decreases with pressure. As the shortwave opacity has less inhomogeneity effect than the IR, we did not specifically treat the visible opacity in this study. For convenience, we fixed the ratio of the visible-to-IR opacity ratio, similar to the analytical framework in \citet{zhangInhomogeneityEffectInhomogeneous2023}. The ratio in our nominal models is unity but we would also test different ratios in Section \ref{sec:alpha}.

\subsection{Hot Jupiter Simulation Setup}
We performed both local and global simulations. The local simulations are carried out to approximate the global-mean situation for comparison. Usually, in 1D models, an averaged stellar incident angle has to be assumed (e.g., \citealt{fortneyPlanetaryRadiiFive2007}). The situation in our study is more straightforward because an analytical solution of the global-mean attenuated stellar flux $F_{\rm v}$ in the pure absorption limit can be obtained with the exponential integral equation (e.g., \citealt{zhangRadiativeForcingStratosphere2013}):
 \begin{equation}
F_{\rm v} (\tau_{\rm v})= -2\sigma T^4_{eq} E_3({\tau_{\rm v}}), \label{lvflx}
 \end{equation}
where $\tau_{\rm v}$ is the optical depth in the visible band. $\sigma$ is the Stefan Boltzmann constant. $E_n(x) = \int_1^{\infty} e^{-wx}w^{-n} dw$ is the exponential integral function. At the TOA, $F_{\rm v} (\tau_{\rm v}=0)= -\sigma T^4_{eq}$ is the globally averaged stellar flux (negative means downward flux). We applied uniform stellar heating in our local simulations based on the above equation. For global simulations, the area scaling is considered to satisfy the energy conservation in the spherical, extended atmosphere. The stellar flux is distributed according to the local incident angles in the visible band on the dayside with no flux on the nightside (see Appendix \ref{app:rte}):
\begin{equation}
F_{\rm v} (\tau_{\rm v}, \mu_0) =
\begin{cases}
-4\mu_0\sigma T^4_{eq} e^{-\tau_{\rm v}/\mu_0}\frac{r_0^2}{r^2}, ~~~~\text{dayside}\\
0. ~~~~~~~~~~~~~~~~~~~~~~~~~~~~\text{nightside}
\end{cases}
\label{glbvflx}
\end{equation}

As discussed in Appendix \ref{app:flx}, using the normalized fluxes to compare the local and global models is more convenient because of the area change with height in the spherical atmosphere. Moreover, due to our primary focus on the ratio of the internal heat flux to the incoming stellar flux, we will \textit{only use the normalized fluxes} in the rest of the paper. In the local model, we normalized the fluxes by $\sigma T_{eq}^4$; while in the global model, we normalized the energy power dividing the flux by $r_0^2 \sigma T_{eq}^4/r^2$. The global average of the normalized incident flux \Cref{glbvflx} is:
\begin{equation}
\overline{F_{\rm v}}=-2 E_3({\tau_{\rm v}}), 
\end{equation}
where $\overline{A}$ denotes the temporally and horizontally averaged quantity of $A$ over the entire domain. The above expression is the same as the normalized mean flux in the local Cartesian model (Equation \ref{lvflx}). HARP also provides flux correction using the Chapman function to modify the attenuated fluxes near the limb (\citealt{kyllingReliableEfficientTwostream1995}). However, we chose not to use it in this study because we ensure the formulation of the global-mean stellar energy attenuation in the 3D global simulation is the same as in the local simulation. 
\begin{deluxetable}{cl}
\tablewidth{0.72\textwidth} 
\tablecaption{Nominal Model Parameters.}
\tablehead{
\colhead{Parameter}  &  \colhead{Value} 
}
\startdata \label{tablepara}
Planetary radius at 200 bar ($R_p$) & $9\times 10^7$ m \\
Rotation rate ($\Omega$) & $2.42 \times 10^{-5}$  s$^{-1}$ \\
Gravity ($g$) & 10 m s$^{-2}$ \\
Specific gas constant ($\mathcal{R}$) & 3777 J kg$^{-1}$ K$^{-1}$ \\
Adiabatic index ($\gamma=c_p/c_v$) & 1.4 \\
Local domain length & $6.4\times 10^6$ m \\
Resolution of 2D local simulations & $64 \times 100$ \\
Resolution of 3D local simulations & $64 \times 64 \times 100$ \\
Resolution of 3D global simulations & $64 \times 128 \times 100$ \\
Temperature-dependent IR opacity &  \Cref{Topa} \\
Temperature-independent IR opacity &  \Cref{TIopa} \\
Visible to IR opacity ratio ($\tau_{\rm v}/\tau_{\rm IR}$) &  1 \\
Temperature at 200 bar ($T_b$)$^{\dagger}$ for $T_{eq}$ = 1200 K & 4300 K\\
\hspace{162pt}1400 K & 6000 K  \\
\hspace{162pt}1600 K & 8100 K  \\
\hspace{162pt}1800 K & 10000 K  \\
\hspace{162pt}2000 K & 10600 K  \\
\hspace{162pt}2200 K & 10500 K \\
\enddata
$^{\dagger}$ The values of $T_b$ are selected for our local models to reproduce the heating efficiencies in \citet{thorngrenBayesianAnalysisHotJupiter2018}.
\end{deluxetable}

Once the model reaches a steady state, we calculate the normalized local internal heat flux ($F_{\rm int}$)--- the net cooling flux from the high-entropy interior---from the TOA flux difference between the IR and visible bands:
\begin{equation}
F_{\rm int}=F_{\rm v,\rm TOA}+F_{\rm IR,\rm TOA}. \label{finteq}
\end{equation}
Because there is no downward IR flux at the TOA, $F_{\rm IR,\rm TOA}$ equals the normalized outgoing longwave radiation (OLR). According to the above definition, $F_{\rm int}$ varies spatially at the TOA but is uniform at the bottom of the model where we have fixed the temperature. One of our goals is to investigate how the uniform $F_{\rm int}$ in the deep convective atmosphere is modulated by the weather layer and exhibits a spatial pattern at the TOA.

The domain-averaged internal heat flux $\overline{F_{\rm int}}$ should be vertically constant throughout the atmosphere. Assuming that the planet is in energy equilibrium, the required heating efficiency ($\eta$) to balance the net cooling from the inflated interior is equal to the normalized $\overline{F_{\rm int}}$ (see Appendix \ref{app:flx}):
\begin{equation}
    \eta=\overline{F_{\rm int}}=F_{\rm IR,\rm TOA}-1.
\label{eff}
\end{equation}
Table \ref{tablepara} lists important parameters in our simulations. We assumed a typical hot Jupiter with a synchronously rotating period of three Earth days (corresponding to the rotation rate of $2.42 \times 10^{-5}$ s$^{-1}$) and gravity of 10 m s$^{-2}$. We assumed a typical hydrogen-dominated atmosphere with the specific gas constant of 3777 J kg$^{-1}$ K$^{-1}$, and the adiabatic index ($\gamma=c_p/c_v$) is assumed to be 1.4 for diatomic gas.

In non-hydrostatic simulations, we use height as the vertical coordinate. We set the bottom pressure at 200 bar, much deeper than the estimated RCBs for inflated hot Jupiters as described in \citet{thorngrenIntrinsicTemperatureRadiativeConvective2019}. We carefully chose the upper boundary height to calculate OLR and internal heat accurately. Due to the significant variation in pressure and temperature in the height coordinate, the optical depth changes dramatically from the dayside to the nightside. We selected the upper boundary height for each case such that the maximum optical depth at the upper boundary (typically near the substellar point) is smaller than 0.01. Considering the temperature is nearly isothermal in the upper atmosphere with grey opacity, neglecting the atmosphere above the upper boundary would generally result in an OLR flux estimate error of less than 0.1\% at the substellar point. The error is much lower at the limbs and on the nightside. Given that total internal heat flux is 1-10\% of the OLR (\citealt{thorngrenBayesianAnalysisHotJupiter2018,sarkisEvidenceThreeMechanisms2021}), the choice of the upper boundary can introduce bias in our estimate on the order of 0.01-0.1\% of the global-mean internal heat flux, which is acceptable for our purpose.

\begin{figure}
\centering \includegraphics[width=0.45\textwidth]{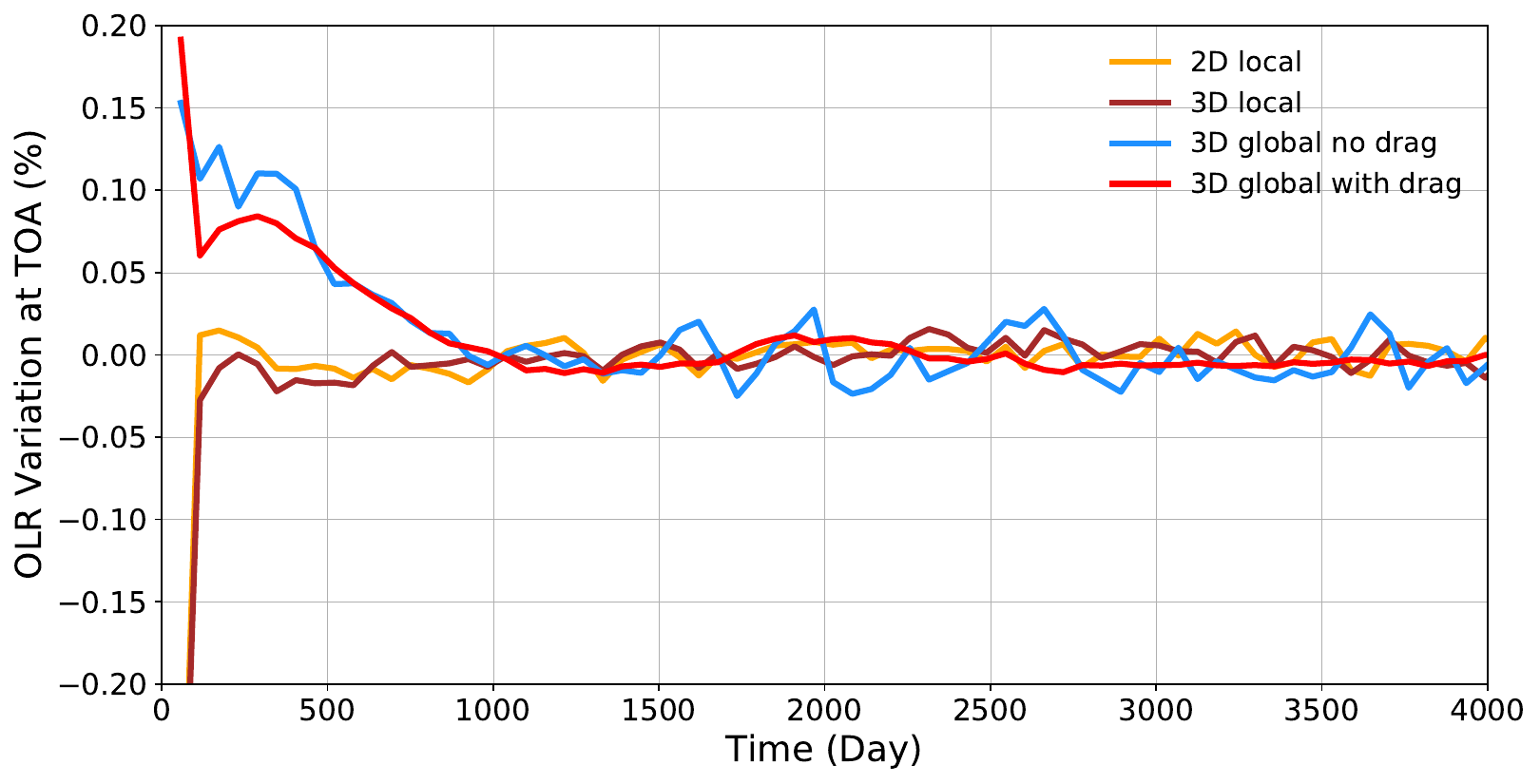}
\caption{Time variation of outgoing longwave Radiation (OLR) at TOA. The variation is defined as the fractional deviation of the OLR normalized by its time-averaged flux after 1500 Earth days of each simulation. Four simulations for $T_{eq}$= 1600 K are shown: 2D (orange) and 3D (brown) local simulations, as well as 3D global simulations without drag (blue) and with a strong drag (red). The systems appear to reach steady states after about 1500 Earth days. The variation is within 0.05\% (i.e., 500 ppm) in a steady state. }\label{convg}
\end{figure}

Most current GCMs define their energy bottom boundary condition using intrinsic flux characterized by the intrinsic temperature $T_{\rm int}$. To analyze the net cooling flux from the interior, we set a reflective bottom boundary in the radiative scheme to prevent any radiative loss to the interior. We then linearly relaxed the bottom boundary temperature to $T_{b}$ to approximate the connection with a constant deep interior entropy. The relaxed temperature is nearly uniform at the bottom boundary with a very short relaxation timescale ($10^3$ s). The underlying assumption is that vigorous convection would homogenize the temperature in the deep convective zone along the isobar. This setup is similar to some recent hot Jupiter simulations (\citealt{tanAtmosphericCirculationUltrahot2019,komacekEffectInteriorHeat2022}).

\begin{figure*}
\centering \includegraphics[width=0.85\textwidth]{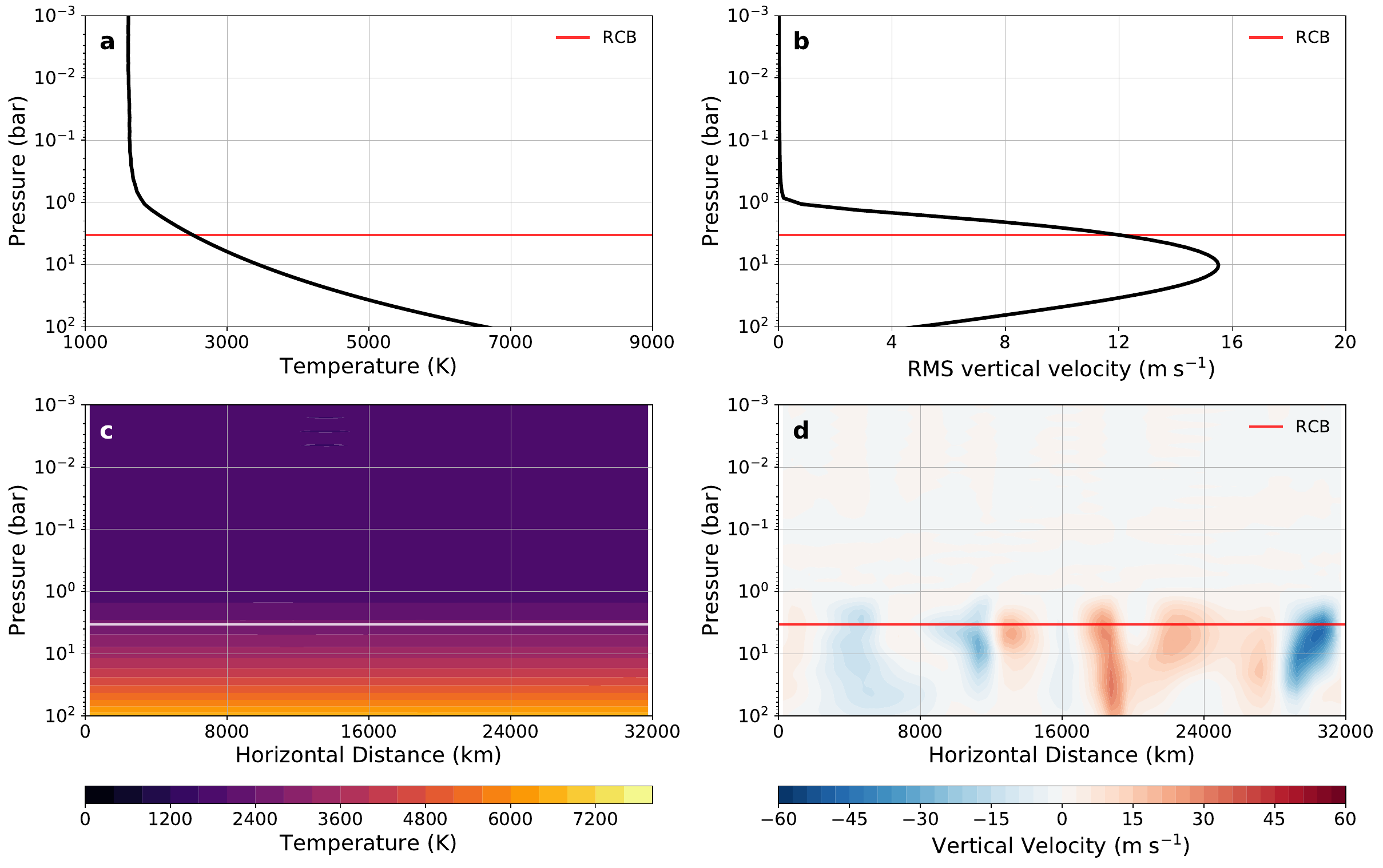}
\caption{Atmospheric structure of a 2D local simulation of a planet of $T_{eq}$= 1600 K. (a): 1D domain-averaged temperature profile. (b): 1D root-mean-square vertical velocity profile. (c): 2D temperature distribution. (d): 2D vertical velocity distribution. The horizontal lines indicate the RCB location ($\sim$3 bars).}\label{localp}
\end{figure*}

By employing local simulations, we meticulously selected the bottom boundary temperature $T_{b}$ for each $T_{eq}$ so that our calculated normalized internal heat flux---the required heating efficiency $\eta$ to keep the energy equilibrium of inflated hot Jupiters (Equation \ref{eff})---in the local model aligns with that from the previous estimates using 1D models (e.g., \citealt{thorngrenBayesianAnalysisHotJupiter2018,sarkisEvidenceThreeMechanisms2021}). The distinction is that our model does not require the convective adjustment scheme adopted in the 1D models. Although local dynamics are present in a small domain, the local simulations are generally spatially homogeneous enough to approximate the global-mean case for our investigation. We then apply the same $T_{b}$ to the global models and analyze the heating efficiency in the global model. 

We implemented a top and bottom sponge layer using a linear damping scheme to dampen the momentum in three dimensions: $\vec F_{\rm spg}=\vec{\bf{u}}/\tau_{\rm spg}$, where $\vec{\bf{u}}$ is the 3D velocity field. The timescale ($\tau_{\rm spg}$ in seconds) is a rapidly decaying function of the distance between a layer and the boundary $\Delta z$ (similar to \citealt{skamarockTimesplitNonhydrostaticAtmospheric2008,mendoncaThreedimensionalCirculationDriving2018}):
\begin{equation}
\tau_{\rm spg} = 10^4\sin^{-2}\left(\frac{\pi(L-\Delta z)}{2L}\right),
\end{equation}
where $0\le\Delta z < L$. $L$ represents the sponge layer width, which is chosen to be approximately five atmospheric layers in our models. Although momentum damping is applied, the energy equation is solved independently, ensuring energy conservation during damping. No additional treatment is needed to convert the damped kinetic energy to thermal energy in our model.

The local domain has a length of $6.4\times 10^6$ m, or roughly 3.6 degrees. We chose 64 evenly-spaced horizontal grids and 100 vertical grids. We tested 2D and 3D local simulations, and their TOA fluxes were consistent. The global simulations also have 100 layers with a horizontal resolution of $64 \times 128$, corresponding to approximately 3 degrees in latitude and longitude. Our sensitivity tests show that the results in this study remain unchanged at higher resolutions (e.g., $96 \times 192$). We performed two global simulations for each $T_{eq}$, with and without drag. The drag case includes a strong drag force $\vec F_{drag}=\vec{\bf{u}}/\tau_{drag}$ throughout the atmosphere, where the drag timescale is $\tau_{drag}=10^4$ s.

Using our fully implicit scheme, we adopted a typical numerical timestep with CFL=5-10, approximately 100-200 seconds. We tested the simulations with smaller time steps (e.g., CFL=3) and ensure our choice of the timestep does not affect the numerical results. Given the large timestep, we updated our radiative scheme at every timestep. We ran all cases for several thousand Earth days until the OLR flux reached a steady state. We also ran some cases for more than 10,000 days to ensure a steady state. The time evolution of the OLR flux fluctuation for typical local and global simulation cases is shown in \Cref{convg}. The amplitude of the flux fluctuation in the drag-free global simulation is well within 0.05\%. The variability in the strong drag case and local simulations is much smaller. Almost all simulations reach a steady state (in terms of OLR) in about 1500 Earth days. We averaged the simulation results for detailed flux analysis over the last 50 Earth days.

Given the large radiative timescale in the deep, optically thick atmosphere, local dynamics may not fully converge in the deep atmosphere. However, the temperature profile in the deep convective zone closely aligns with the adiabat due to efficient convective heat transport below the RCB. The adiabatic temperature profile starts from the bottom boundary where the temperature is quickly relaxed back to a fixed value. Once the temperature profile reaches a steady state, the radiative flux at the TOA remains unchanged in our simulations. Our modeling approach is similar to that in \citet{tanAtmosphericCirculationUltrahot2019} and \citet{komacekEffectInteriorHeat2022}, contrasting with other hot Jupiter GCMs that incorporate bottom heat flux. In those models, reaching a steady state in the deep convective zone  can take a very long time (e.g., \citealt{schneiderNoEvidenceRadius2022,sainsbury-martinezEvidenceRadiusInflation2023}).

In Sections \ref{sec:local} and \ref{sec:global}, we will demonstrate the difference in heat flow transport between the local and global models using our nominal simulations with $T_{eq}=1600$ K. We selected this temperature because the heating efficiencies of inflated hot Jupiters derived from two previous studies (\citealt{thorngrenBayesianAnalysisHotJupiter2018,sarkisEvidenceThreeMechanisms2021}) are roughly comparable (about 2.3\%, see Figure 6 in \citealt{sarkisEvidenceThreeMechanisms2021}). Additionally, $T_{eq}=1600$ K is located around the predicted maximum heating efficiency from the Ohmic heating mechanism (e.g.,\citealt{batyginEvolutionOhmicallyHeated2011,rogersMagneticEffectsHot2014,ginzburgExtendedHeatDeposition2016}).

We set the bottom temperature $T_{b}$ to 8100 K in all simulations, allowing the local simulations to yield a normalized internal heat flux of about 2.3\% for $T_{eq}=1600$ K, consistent with the previous 1D model results. It is worth noting that the value of $T_{b}$ is not necessarily equal to the 1D models with the non-grey opacity. However, since our local model and global models use the same opacity scheme, we can still compare them to investigate the difference in interior cooling in the local and global simulations and the underlying mechanisms.

\section{Local Simulations with $T_{eq}=1600$ K} \label{sec:local}

We first present local simulations with $T_{eq}=1600$ K to demonstrate their typical atmospheric structure and energy flow patterns. The 2D and 3D local models are consistent in domain-averaged temperature and vertical wind speed. Thus we present only 2D simulations as a proxy for the global mean scenario. \Cref{localp} illustrates the typical temperature and vertical wind distributions. The temperature distribution is horizontally homogeneous, decreasing from 7500 K at the domain's bottom to approximately 1400 K in the upper atmosphere, where the temperature is nearly isothermal. The vertical wind distribution exhibits temporal and spatial variabilities in the lower atmosphere but lacks well-organized convective cells in the convective zone. The maximum upward and downward velocities reach several tens of $\rm{ms^{-1}}$ in the convective zone. The root mean square (RMS) of the vertical velocity distribution increases from 5 $\rm{ms^{-1}}$ at 100 bar to about 16 $\rm{m~s^{-1}}$ at 10 bars, then rapidly decreases towards zero in the top radiative zone, indicating the quick suppression of vertical motion in the atmosphere when convection ceases.

\begin{figure*}
\centering \includegraphics[width=0.95\textwidth]{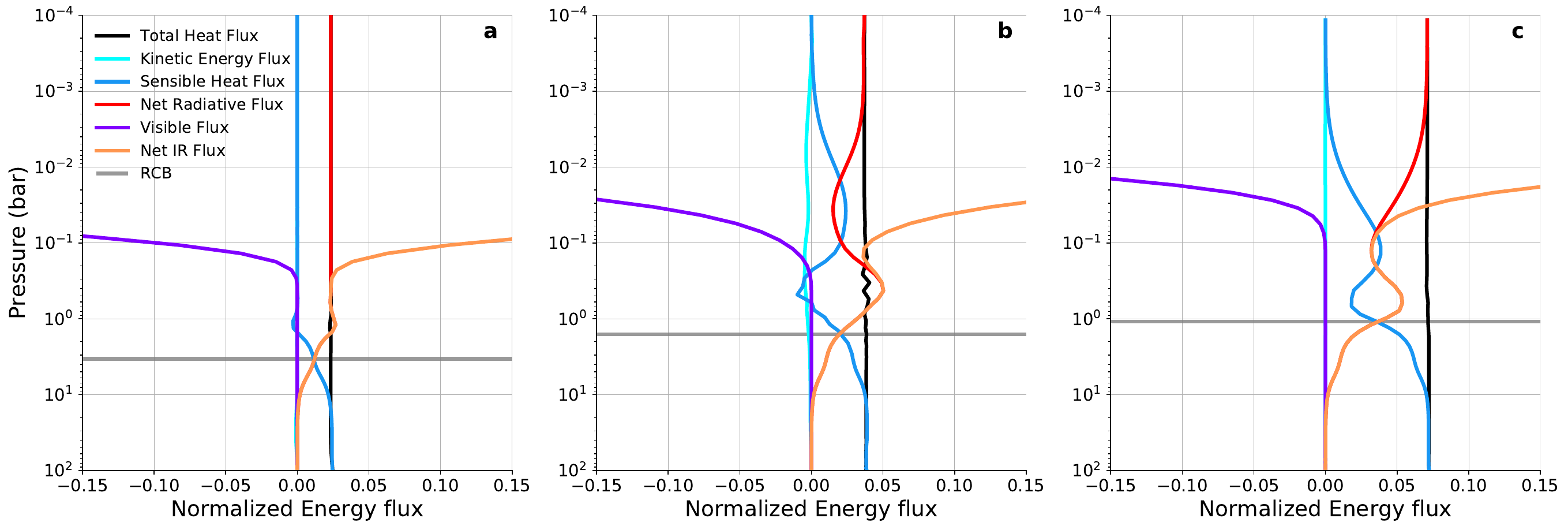}
\caption{Normalized domain-averaged energy fluxes in (a) the 2D local simulations and (b) 3D global simulations without drag and (c) with strong drag for $T_{eq}$= 1600 K. The fluxes are normalized by the global-mean stellar irradiation. We define upward flux as positive. The pressure on the y-axis is the global-mean pressure at that altitude, as we use the altitude coordinate in the non-hydrostatic simulations. Several mean fluxes (see Equation \ref{vflxeq}) are shown: radiative flux in the visible band (purple), IR band (orange), net IR flux (orange), net radiative flux (red), sensible heat flux (blue), the vertical flux of the kinetic energy (cyan), and the total heat flux (black). The total heat flux is very close to a straight line, indicating excellent energy conservation in our model. The small deviation comes from the time variability that was not entirely averaged out in the last 50 Earth days. The horizontal lines define the deepest intersections of the sensible heat flux and the net radiative flux, which mark an approximate location of the RCB in the global-mean sense. This plot only shows the fluxes from 100 bar to 0.1 mbar, above which the fluxes remain constant to the TOA (smaller than 1 Pa).}\label{eflx}
\end{figure*}

We calculate the normalized domain-averaged vertical energy fluxes to analyze the heat flow. The mean vertical flux equation in a steady state is derived in Appendix \ref{app:flx} (Equation \ref{flx-eq}):
\begin{equation}
\overline{\rho c_p wT} + \overline{\rho w E_k} + \overline{F_{\rm v}}+\overline{F_{\rm IR}}=\overline{F_{\rm int}}, \label{vflxeq}
\end{equation}
where $\rho$ is the density, $T$ is temperature, and $c_p$ is the specific heat at the constant pressure. $F_{\rm v}$ and $F_{\rm IR}$ are the net radiative fluxes in the visible and infrared bands, respectively. In local 2D models, the specific kinetic energy $E_k = \frac{1}{2}(u^2 + w^2)$ where $u$ and $w$ are the velocities in horizontal ($x$) and vertical ($z$) directions, respectively.

The left-hand side of \Cref{vflxeq} represents heat transport by atmospheric dynamics such as convection, waves, and large-scale circulation. The first term is the vertical sensible heat flux $\rho c_p wT$, approximately on the order of $\rho w c_s^2$ where $c_s$ is the sound speed $c_s=(\gamma c_pT)^{1/2}$. The second term is the vertical flux of kinetic energy $\rho w E_k$. In our 2D local simulations, the velocities $u$ and $w$ are generally much smaller than the sound speed. Consequently, the kinetic energy flux term is negligible compared with the sensible heat flux. The first two terms of the right-hand side are the radiative fluxes from the visible and IR bands, respectively.

\begin{figure*}
\centering \includegraphics[width=0.85\textwidth]{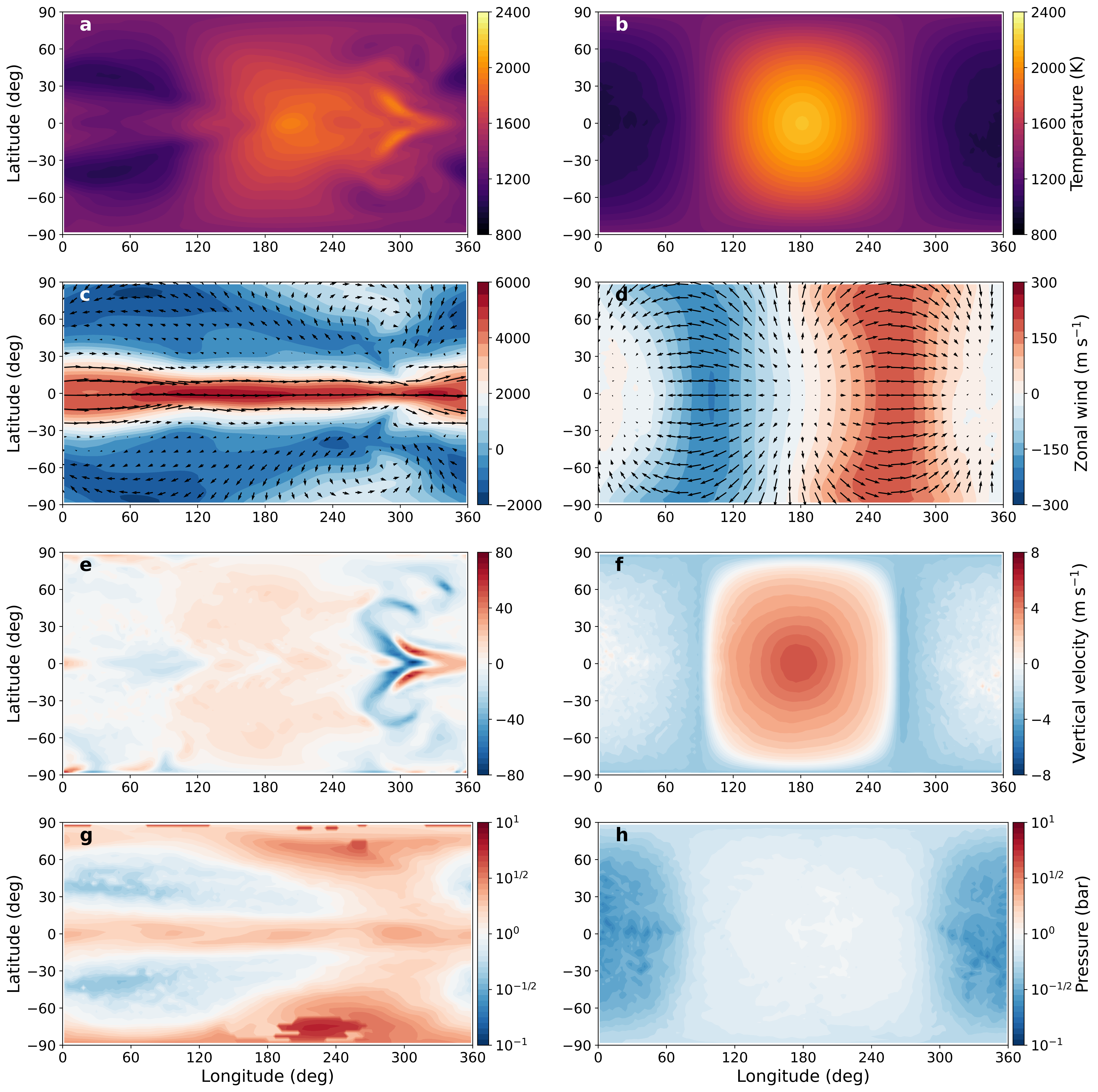}
\caption{Horizontal atmospheric structure of 3D global simulations of $T_{eq}$= 1600 K without (left) and with strong drag (right column). First row: temperature map at 0.1 bar. Second row: east-west wind map at 0.1 bar, with wind direction indicated by color and arrows representing the magnitude. Third row: vertical velocity map at 0.1 bar. Fourth row: spatial map of the local RCB across the globe.
}\label{latlon2d}
\end{figure*}
The net cooling flux from the interior---the internal heat flux $\overline{F_{\rm int}}$---is a residue from the system's imbalance of dynamical and radiative heat fluxes. Due to energy conservation, the internal heat flux should be vertically constant in the 2D simulation. This can be used as a sanity check for the model. The sensible heat flux, kinetic energy flux, and radiative fluxes in \Cref{vflxeq} are plotted in \Cref{localp}. The sum of the net energy flux is nearly constant throughout the atmosphere, indicating good energy conservation in our model.

The interaction between radiation and convection determines the heat transport regimes in the local simulations. As the vertical transport of kinetic energy is negligible, we define the location of the RCB using the intersection between the sensible heat flux and the net radiative flux (\Cref{eflx}a). In this case, the RCB is approximately at 3 bars, consistent with the argument in \citet{thorngrenIntrinsicTemperatureRadiativeConvective2019} that a high-entropy hot Jupiter should have a shallow RCB (also see \citealt{sarkisEvidenceThreeMechanisms2021,komacekEffectInteriorHeat2022}). 

The internal heat flux is transported as sensible heat flux in the lower atmosphere. The flux is almost constant with pressure as it is carried out by convection but decreases quickly above the RCB. This is consistent with the vertical velocity distribution in \Cref{localp}, where the velocity quickly drops above the RCB. The downward (negative) stellar flux decreases from the top atmosphere to near zero at about 0.5 bar where the optical depth is large. In contrast, the net infrared radiation (upward plus downward) is negligible in the convective zone. But it increases quickly above the RCB and becomes larger than the sensible heat flux. The total net radiative flux in both visible and IR bands dominates heat transport above the RCB and becomes constant with pressure in the upper radiative zone.

The rapid decay of the sensible heat flux and the dominance of radiation above the RCB suggest that the local simulations are generally in RCE. Thus, we can use them to mimic traditional 1D RCE models that assume full heat distribution across the globe. The primary differences between our local models and previous 1D models are that we used the globally averaged incoming stellar flux instead of choosing a ``global-mean" incident angle and directly simulated the radiative-convective interaction rather than using a convective-adjustment scheme. With parallel computation, the local model is fast and may be useful as an alternative RCE model to study the first-order, global-mean properties of the atmosphere. However, it should be used cautiously, as the difference between the local and 3D global models can sometimes be significant, as demonstrated later.

\section{Global Simulations with $T_{eq}=1600$ K} \label{sec:global}

We performed two 3D global simulations with $T_{eq}$= 1600 K. The first one has no wind drag except in the sponge layers, and the second one includes a strong drag of $\tau_{drag}=10^4 $ s throughout the atmosphere. These two cases represent the extremes with the highest and lowest heat redistribution efficiencies in the atmosphere. By comparing the two cases and the global versus local simulations, we can investigate the roles of 3D geometry, radiation, and dynamical processes on atmospheric inhomogeneity and heat flow transport. We organize the results into four subsections: weather pattern, global-mean flux, RCB morphology, and the TOA flux.

\subsection{Weather Pattern}

The two global simulations yield very different weather patterns and spatial homogeneities in the atmosphere, as shown in \Cref{latlon2d}. We illustrate the differences in the global maps of temperature, horizontal wind, and vertical velocity at a pressure level of 0.1 bar, which probes the stratified region overlying the convective zone.

The drag-free case exhibits a typical weather pattern of hot Jupiter atmospheres, similar to those in Figure 9 in \citet{hengAtmosphericDynamicsHot2015}, the ``nominal hot Jupiter case" in Figure 14 of \citet{zhangAtmosphericRegimesTrends2020}, and recent simulations with hot interiors in \citet{komacekEffectInteriorHeat2022}. The horizontal temperature map (\Cref{latlon2d}a) is primarily shaped by the eastward zonal wind and eastward-propagating Kelvin wave at the equator and the westward-propagating Rossby modes in the off-equatorial flanks. The equatorial superrotating jet reaches speeds as high as 6000 $\rm m~s^{-1}$ at this level and covers latitudes within $\pm$ 30 degrees (\Cref{latlon2d}c). The vertical wind flow (\Cref{latlon2d}e) is generally upwelling on the dayside and downwelling on the nightside at 0.1 bar, except in the equatorial region between 270 and 320 degrees in longitude. In the region slightly eastward of the evening terminator, strong upwelling and downwelling plumes are observed, with vertical velocities reaching 80 $\rm m~s^{-1}$. This area behaves like sinks and chimneys in hot Jupiter's atmosphere. It promotes intense tracer mixing between the lower and upper atmospheric layers (see Figure 3 in \citealt{parmentier3DMixingHot2013} and Figure 2 in \citealt{zhangGlobalmeanVerticalTracer2018a}).

\begin{figure*}
\centering \includegraphics[width=0.7\textwidth]{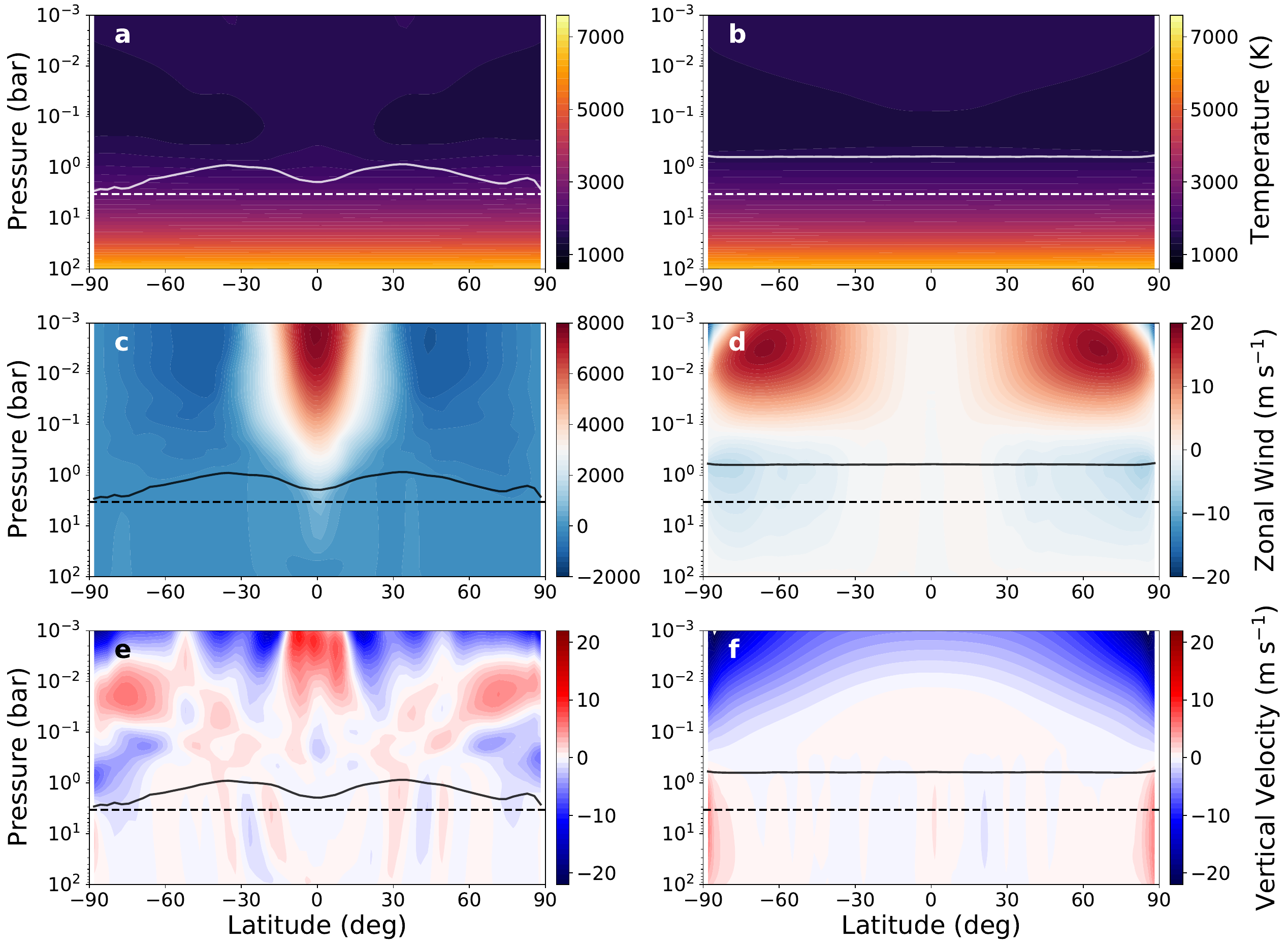}
\caption{Latitude-pressure distributions of zonal-mean temperature (upper), zonal-mean zonal wind (middle), zonal-mean vertical wind (bottom) from the global simulations of $T_{eq}$= 1600 K without (left) and with a strong drag (right column). The horizontal dash line marks the location of the RCB in the 2D local simulations. The solid lines indicate the local RCBs across the substellar point in the 3D global simulations.}\label{latp2d}
\end{figure*}

\begin{figure*}
\centering \includegraphics[width=0.7\textwidth]{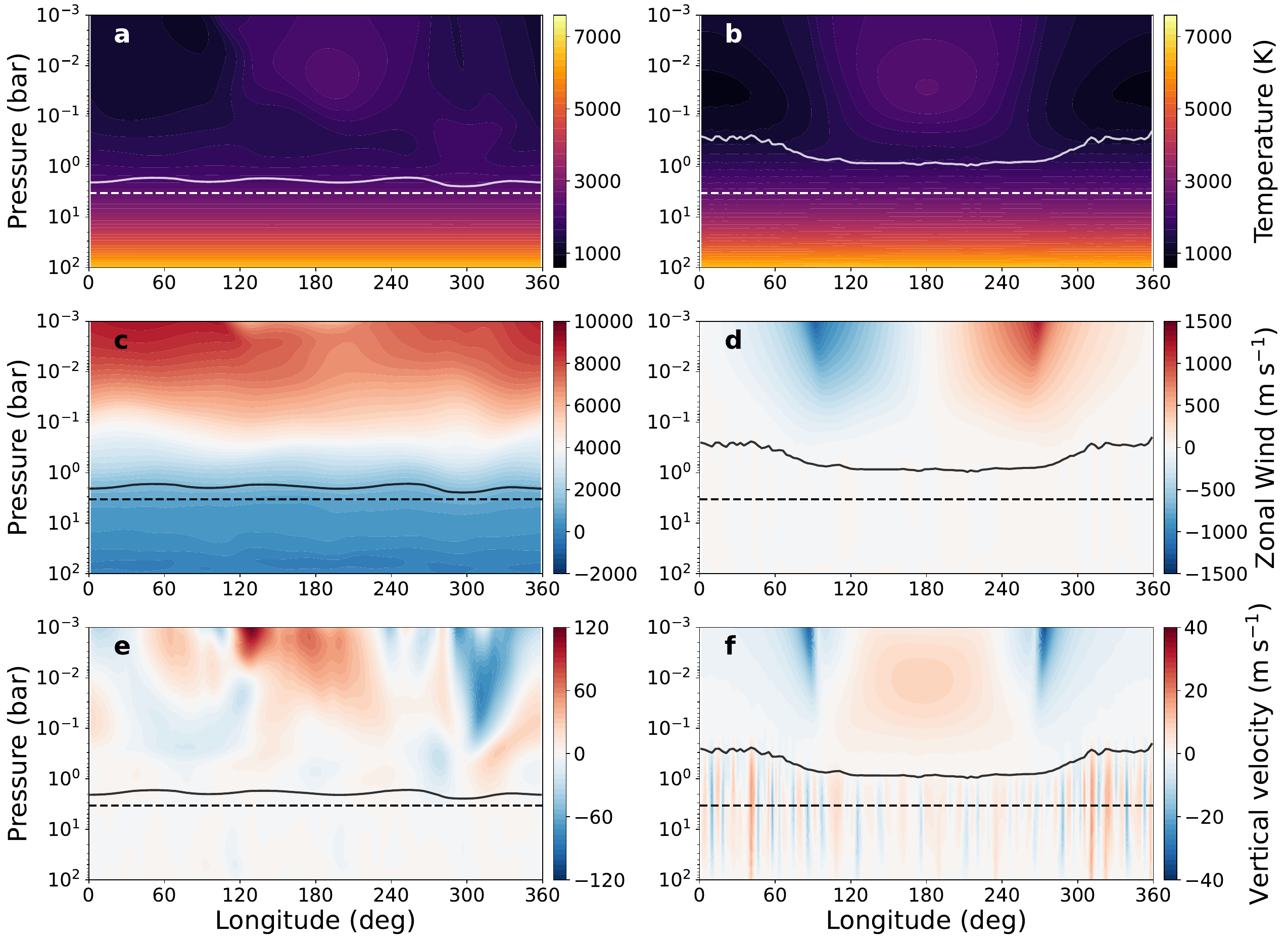}
\caption{Similar to \Cref{latp2d} but for the longitude-pressure distributions.}\label{lonp2d}
\end{figure*}

Compared to the drag-free case, the horizontal structure of the simulations with a strong drag appears relatively simple. A pronounced day-night temperature contrast ranges from approximately 2400 K at the substellar point to 800 K on the nightside (\Cref{latlon2d}b). The temperature distribution generally follows the stellar incident angles. The formation of zonal jets is suppressed by the strong drag. Horizontal wind speed is also significantly reduced, displaying a divergent pattern from the substellar to the anti-stellar point. The maximum wind speed reaches only about 200 $\rm m\;s^{-1}$ (\Cref{latlon2d}d), much slower than the zonal jet speed in the drag-free case. The vertical wind distribution exhibits a strong upwelling pattern on the dayside and a broad, yet weaker downwelling on the nightside, with wind speeds of a few $\rm m\;s^{-1}$. The weather pattern in the strong drag case aligns with previous simulations with strong drags in hydrostatic models (e.g., \citealt{komacekVerticalTracerMixing2019}).

The horizontal weather patterns in the stratified region can also be understood within the framework of Helmholtz decomposition (\citealt{hammondRotationalDivergentComponents2021,lewisTemperatureStructuresAssociated2022}). The horizontal circulation can be decomposed into divergent (``vorticity-free") and rotational (``divergence-free") components. In the drag-free simulations, the circulation can be dominated by a substellar-to-anti-stellar divergent flow and a rotational flow composed of a zonal-mean zonal jet and a stationary eddy wave pattern with wavenumber-1. However, the strong-drag simulation is primarily dominated by the divergent pattern, as the rotational jet and waves are substantially suppressed by the strong drag and radiation.

To illustrate the distinct vertical structures between the two simulations, we further present the zonally averaged temperature, zonal wind, and vertical velocity in the latitude-pressure plane (\Cref{latp2d}), as well as a vertical slice of these variables at the equator as functions of longitude and pressure in (\Cref{lonp2d}). Both the drag-free and strong drag cases exhibit relatively similar zonal-mean temperature patterns (\Cref{latp2d}a and b), with an adiabatic structure in the convective zone below and a temperature inversion in the stratified region at around 0.1 bar. The temperature minimum in the stratified zone appears thicker and extends higher at high latitudes. However, the equatorial slices of the temperature distribution from the two cases differ. In the drag-free case, the vertical tilt of the temperature pattern at the equator (\Cref{lonp2d}) suggests that the temperature structure is more influenced by the zonal wind and waves at lower altitudes. In contrast, the temperature is consistently hotter at the substellar point and decreases towards the nightside at the equator in the strong drag case.

The zonal-mean zonal wind in the drag-free case develops a coherent vertical super-rotating structure at the equator in the stratified zone, accompanied by two westward jets at mid-latitudes. The equatorial jet is suppressed above the convective zone. In this case, we did not observe the formation of multiple jet patterns in the convective region. The horizontal wind pattern is divergent in the strong drag case (\Cref{latlon2d}d). The eastward and westward components largely cancel each other out in the zonal-mean zonal wind, with a maximum value of about 20 $\rm m~s^{-1}$ at high latitudes (\Cref{latp2d}d). In the longitude-pressure plane at the equator, the zonal wind distribution in the drag-free case displays a clear wave pattern along the longitude (\Cref{lonp2d}c), while the strong-drag case exhibits a divergent wind pattern around the substellar point (\Cref{lonp2d}d). The zonal-mean vertical wind is weak in both cases (\Cref{latp2d}e and f), but the equatorial vertical wind is stronger (\Cref{lonp2d}e and f). The strong drag case reveals a distinct pattern of multiple upwelling and downwelling plumes in the convective zone that transition into a broad upward flow on the dayside in the stratified zone (\Cref{lonp2d}f). These plumes do not seem to be present in traditional hydrostatic GCMs that solve primitive equations (private communication with Thaddeus Komacek). This suggests that how efficiently mixing happens in the deep convective atmosphere might vary between hydrostatic and non-hydrostatic models.

\subsection{Global-mean Vertical Energy Flux}

The global mean vertical energy fluxes in the 3D global simulations are calculated slightly differently from the local case because we must consider the surface area change with altitude in an extended, spherical atmosphere. In 3D spherical geometry, the vertically conserved quantity is no longer the heat flux, but the energy power---the integrated flux over the global area. However, with the normalization proposed in Appendix \ref{app:flx}, we can achieve the same formalism of the flux equation as in the local model (Equation \ref{vflxeq}). In the 3D simulations, the specific kinetic energy $E_k = \frac{1}{2}(u^2 + v^2 +w^2)$, where $u$, $v$, and $w$ represent the azimuthal (east-west $\phi$), polar (north-south $\theta$), and radial (vertical $r$) directions in spherical polar coordinates.

Similar to the local simulations, the normalized internal heat flux, which essentially represents the scaled internal luminosity (see Appendix \ref{app:flx}), remains constant throughout the atmosphere (\Cref{eflx}). This suggests good energy conservation in our global simulations. However, the global-mean stellar flux penetrates less deeply in the global models than in the local models (\Cref{eflx}). This is due to the temperature-dependent opacity, where the dayside temperature in the global simulations exceeds 2000 K, much hotter than the temperature at the same pressure level in the local models (\Cref{latlon2d}ab). As a result, the dayside opacity increases dramatically (\Cref{opa}), leading to a larger absorption of the incoming stellar flux than in the local model. The impact of opacity will be further discussed in Section \ref{sec:opa}.

Like the local simulations, the vertical flux of the kinetic energy is negligible in the global simulations. This is mainly due to the wind velocity being much less than the sound speed, which is generally true for the strong drag case where drag damps the winds. This is also true in the convective region in the drag-free global simulations. However, in the stratified region, the zonal wind velocity is close to the sound speed (\Cref{latlon2d}c). The zonal-mean zonal wind can even reach as fast as 8000 $\rm m~s^{-1}$ in the upper atmosphere (\Cref{latp2d}c). Nevertheless, the global-mean kinetic energy flux ($\overline{\rho wE_k}$) depends less on the wind speed magnitude but critically on the correlation between the vertical wind pattern and the zonal wind pattern. As illustrated in \Cref{latp2d}c, the zonal jet speed is approximately uniform at the equator. Meanwhile, the vertical wind shows strong upwelling (positive) on the dayside and downwelling (negative) on the nightside (\Cref{latp2d}e). Therefore, the upward dayside kinetic energy flux is almost balanced by the downward flux on the nightside, resulting in a small net flux. On the other hand, the vertical velocity pattern is correlated much better with the temperature pattern on both day and nightsides (\Cref{latp2d}a and e), including the chevron pattern near the evening terminator. The upward transport of the hotter air on the dayside and the downward transport of the colder air on the nightside contribute to the net sensible heat flux in the global-mean sense. This idea of considering temperature as a tracer in the above analysis is similar to the method of studying global-mean vertical tracer mixing in planetary atmospheres (\citealt{holtonDynamicallyBasedTransport1986,zhangGlobalmeanVerticalTracer2018a,zhangGlobalmeanVerticalTracer2018}).

The sensible heat and net radiative flux distributions in 3D global simulation differ dramatically from the local simulations. In global cases, the two fluxes intersect multiple times. Starting from the bottom of the model, the sensible heat flux dominates the convective heat transport, and radiation is inefficient in the optically thick region below. This is evident from the strong upward and downward plumes at the equator in the strong-drag case (\Cref{lonp2d}). The two fluxes intersect at about 1.5 bar in the drag-free case and 1 bar in the strong-drag case (\Cref{eflx}). This first intersection (counted from below) marks the RCB of the atmosphere in the global mean sense.

Above the RCB, the sensible heat flux quickly decreases towards 0.5 bar in both cases (\Cref{eflx}). In the drag-free case, the sensible heat flux can even become less than zero, implying a downward transport of heat by dynamics. Although this behavior appears similar to that in the local models, the sensible heat flux in both global simulations quickly turns around and becomes larger above approximately 0.5 bar. At the same time, the net radiative heat flux increases above the RCB and maximizes at 0.5 bar, then decreases towards 0.1 bar and intersects with the sensible heat flux again. Above this intersection level, the sensible heat flux quickly drops to zero, and radiation dominates the heat transport to the top of the atmosphere.

Traditionally, the vertical structure of a planetary atmosphere is divided into two zones: an underlying convective zone and an overlying radiative zone. This is true as shown in our local models (\Cref{localp}). However, the behavior in the global-mean vertical fluxes from the 3D global simulations indicates that the atmosphere of 3D hot Jupiters is much more complex. The lowest atmosphere is still convective. But the atmosphere above the RCB cannot be simply regarded as a single radiative zone. This part of the atmosphere is generally stratified (\Cref{latp2d}a and b), so the heat cannot be transported by convection. However, the large-scale general circulation plays a crucial role in transporting the heat upward. Thus, we may refer to the overlying atmosphere above the RCB as the ``radiation-circulation" zone on hot Jupiters. 

The multiple intersections between the sensible heat flux and the net radiative flux indicates that the ``radiation-circulation" zone might be separated into several regions with different dominant heat transport processes. In the first region above the RCB, the stratification quickly damps the convection and decreases the sensible heat transport. The general circulation also tends to transport the heat downward, which will be detailed in Section \ref{sec:dyn}. Radiation becomes strong above the RCB with a large temperature gradient and upward IR radiative flux.

Above this first region, both the temperature and vertical wind patterns are shaped by the waves and jets, forming a similar spatial pattern. This coherent pattern leads to a large-scale transport of hotter air on the dayside and colder air on the nightside. At the same region, the inversion in the vertical temperature structure (\Cref{latp2d}a and b) is largely due to absorption of the stellar flux on the dayside, standing waves in the equatorial region, and large-scale jet transport of heat from the dayside to the nightside. This inversion structure causes a substantial downward IR radiative flux above the RCB between 0.5 and 0.1 bar, largely compensating for the upward emitted flux from the hotter atmosphere below. In this region, the atmosphere organizes itself to transport the heat upward through dynamics rather than radiation.

Finally, in the third region above 0.1 bar, the upper atmosphere is optically thin and the radiation becomes very efficient to cool to space, while the dynamical heat flux quickly drops due to the exponentially decreasing density towards the top of the atmosphere.

\subsection{RCB Morphology}

In the global-mean sense, the approximate location of the RCB in the 3D simulations can be estimated by the intersection between the global-mean sensible heat flux and the net radiative flux. As shown in \Cref{eflx}, the location of the ``global-mean" RCB in the drag-free case is higher, at around 1.5 bar, compared to the local 2D simulations (3 bars) as shown in \Cref{eflx}. The RCB is even higher in the strong-drag case, at around 1 bar. These values are generally consistent with the previous 1D models with more realistic opacity {\citealt{thorngrenBayesianAnalysisHotJupiter2018,sarkisEvidenceThreeMechanisms2021}). 

To further illustrate the details, we estimated the local RCB distribution. Although the local RCB is ill-defined due to the impact of horizontal dynamics on heat transport, it serves a useful purpose in tracing the upper surface of the deep convective atmosphere. However, one cannot rely on the local sensible heat flux and the net radiative flux to estimate the local RCB because these two fluxes do not consistently intersect at all locations. To identify the local RCB, we applied the buoyancy frequency (also known as the Brunt-V{\"a}is{\"a}l{\"a} frequency, $N$), which is based on the vertical gradient of the potential temperature ($\theta$):
\begin{equation}
N^2=\frac{g}{\theta}\frac{d\theta}{dz}.
\end{equation}

Our process for mapping the local RCB is as follows: Firstly, we calculated a reference $N^2$ value at the pressure level of the ``global-mean" RCB where the global-mean sensible heat flux and the net radiative flux intersect, based on the global-mean $\theta$ profile. Next, for each location on the sphere, the local RCB was identified as the local pressure level where the local $N^2$ matched the reference value. In situations where the local $\theta$ profiles demonstrate significant variations with altitude, using the $N^2$ value at the ``global-mean" RCB as a reference can often result in a noisy local RCB map. In such cases, we employed $N^2=2\times 10^{-7} s^{-2}$ as the reference value to trace the local RCB.

The derived RCB morphology is illustrated in \Cref{latlon2d} and h for the drag-free and strong-drag cases, respectively. The RCB surfaces in both cases show significant variations in longitude and latitude, which is different from the previous assumption of a longitudinally-uniform and latitudinally-varying RCB (\citealt{rauscherINFLUENCEDIFFERENTIALIRRADIATION2014}). In the drag-free case, the RCB is located deeper in the atmosphere at both the equator and high latitudes, reaching about 10 bars, while in the middle latitudes, the RCB is located higher (above 1 bar). The latitudinal distribution of the RCB is clearly shown in the zonal-mean plots (\Cref{latp2d}), with two bumps in the mid-latitudes. The RCB is also deeper in the eastern hemisphere (evening side) than the western hemisphere (morning side).
\begin{figure*}
  \centering \includegraphics[width=0.85\textwidth]{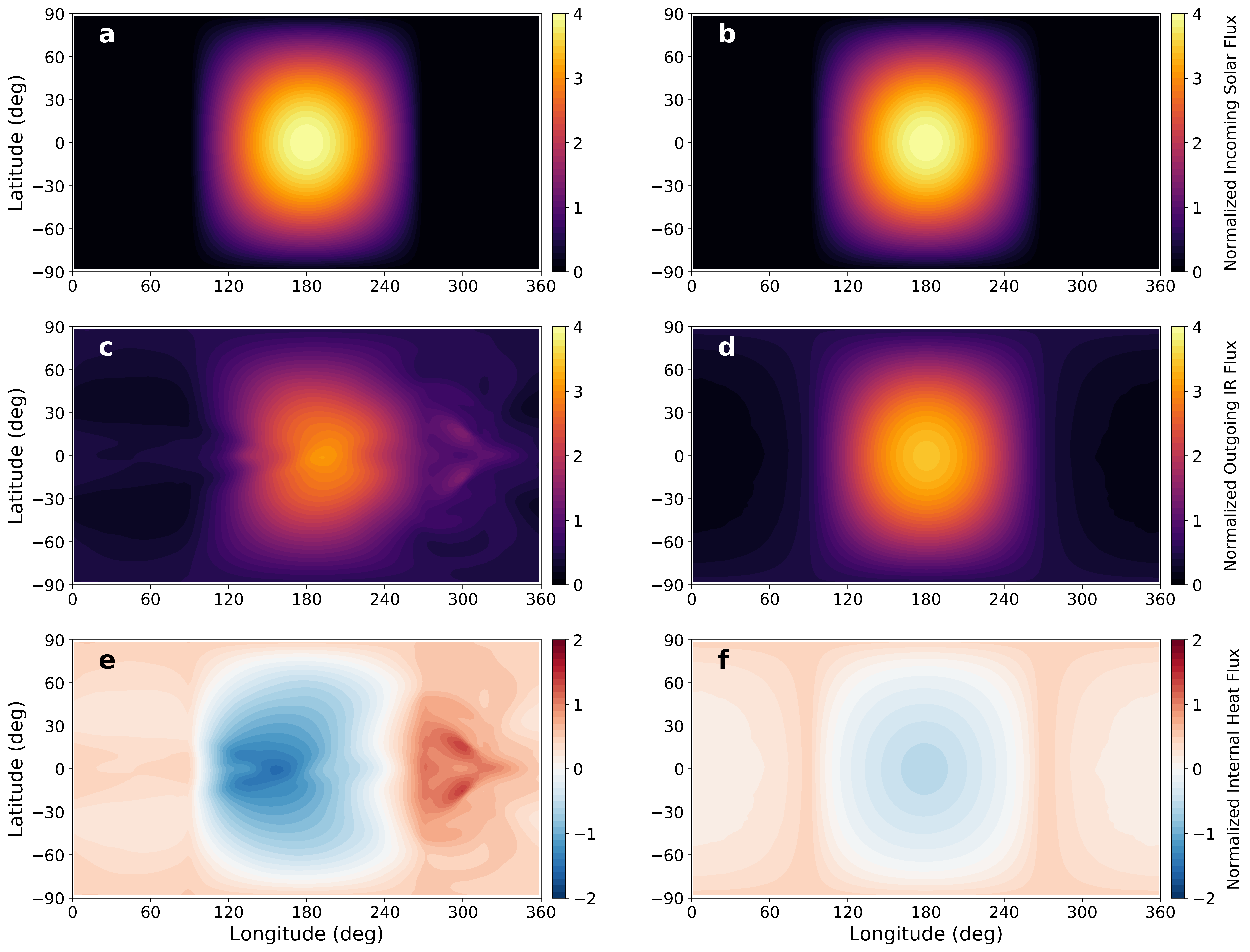} 
  \caption{Spatial distribution of normalized incoming stellar (top), outgoing IR (middle) and net outgoing internal heat fluxes (bottom) from 3D global simulations of $T_{eq}$= 1600 K without (left) and with a strong drag (right column). The fluxes are normalized by the global-mean stellar flux.}\label{toaflx} 
\end{figure*}

Although the local RCB traces the buoyancy frequency associated with the vertical gradient of the potential temperature, the general pattern of the RCB in the drag-free case does not resemble the temperature pattern at 0.1 bar. This implies that the temperature pattern is not vertically coherent from the RCB to the second region in the radiation-circulation zone and is significantly modulated by atmospheric dynamics. The general pattern of the RCB resembles the zonal wind distribution rather than the zonal-mean temperature contour. The correlation of the temperature pattern and vertical velocity determines vertical heat transport, indicating the importance of circulation in the global-mean vertical heat transport above the RCB. \citet{hammondRotationalDivergentComponents2021} has illustrated how dynamics shape the horizontal heat transport from the dayside to the nightside in the radiative zone. The correlation between temperature and vertical velocity patterns determines vertical heat transport, demonstrating the importance of circulation in the global-mean vertical heat transport above the RCB. 

In the strong-drag case, the RCB distribution in \Cref{latlon2d}h does not resemble the horizontal wind pattern but rather correlates with the vertical wind and temperature distributions. The pattern seems more uniformly distributed on the dayside than the temperature and vertical wind patterns as a result of horizontal heat transport. It appears that the heat transport by the rotational component of the horizontal wind is not important. Instead, the divergent flow transports heat from the dayside to the nightside and, together with the vertical heat transport, regulates the morphology of the temperature and RCB. The RCB is deeper on the dayside (a bit higher than the 1 bar level) due to stellar irradiation and lower on the nightside (on the order of 0.1 bar). The RCB is flat with latitude (\Cref{latp2d}). The longitudinal distribution of RCB across the equator clearly separates the stratified region and the underlying convective region, almost following exactly the top of the convective plumes that extend from 1 bar from the dayside to about 0.3 bar on the nightside (\Cref{lonp2d}f).

\subsection{TOA Heat Flux}

The normalized TOA fluxes of the two global cases are shown in \Cref{toaflx}. The incoming stellar flux is concentrated on the dayside, while the outgoing IR flux widely spreads across the globe. The incoming stellar energy distributions in the two cases are the same, with a maximum (normalized) value of 4 at the substellar point, decreasing following the cosine of the incident angle. The outgoing longwave radiation primarily follows the temperature pattern in the stratified zone (\Cref{toaflx}c and d). The difference between the IR flux and the visible stellar flux shows the spatial distribution of the internal heat flux at the TOA. The internal heat flux is uniform in the deep convective atmosphere where we have fixed the bottom temperature. As it is transported upward through the weather layer above the RCB, the flux is largely redistributed by the dynamics and radiation, leading to the observed TOA pattern in \Cref{toaflx}.

The TOA internal heat flux is mostly negative on the dayside, where the internal heat escape is inhibited by the excess stellar energy compared to IR radiation. The flux is positive on the nightside, where the interior cooling is efficient. This is consistent with the argument in \citet{guillotEvolution51Pegasus2002} that the steeper temperature profile on the nightside would facilitate the interior cooling and the ``radiator fin" analogy for hot Jupiter cooling in \citet{zhangInhomogeneityEffectII2023}. For the drag-free case, the most efficient interior cooling occurs around the evening terminator due to the strong vertical heat transport and negligible downward stellar flux. Although the maximum value of the TOA internal heat flux can reach twice the global-mean incoming stellar flux, the upward and downward components largely cancel out, and the global-mean residue interior flux is only a few percent in both global simulations (\Cref{eflx}).

To understand the spatial pattern of the internal heat flux at the TOA, we utilize the vertically integrated energy equation given in Appendix \ref{app:flx}:
\begin{equation}
\langle\nabla_h \cdot \rho(c_p T + \Phi + E_k) \mathbf{u}\rangle + F_{\rm v,\rm TOA} + F_{\rm IR,\rm TOA}-\overline{F_{\rm int}}=0,
\end{equation}
where $\langle\cdot\rangle$ denotes the vertical integration along the radial direction. Here, $c_p T + \Phi$ represents the dry static energy and $\Phi$ is the geopotential (see Appendix \ref{app:flx}). The inclusion of kinetic energy and internal heat flux in this equation, which differs from the traditional framework under primitive equations such as Equation (8) in \citet{hammondRotationalDivergentComponents2021}, ensures energy conservation and eliminates the need for discussions on energy conversion from potential to kinetic energy. However, the physical essence remains the same.

The vertically integrated equation does not aid in the understanding of the vertical transport of the internal heat flux as $\overline{F_{\rm int}}$ is considered as a background. Nevertheless, it provides valuable insights into the horizontal pattern of TOA $F_{\rm int}$ flux, which should be anti-correlated with the divergence of the horizontal energy flux term $\langle\nabla_h \cdot \rho(c_p T + \Phi + E_k) \mathbf{u}\rangle$. If we assume the ``Weak Temperature Gradient" approximation (which we may refer to as the ``Weak Energy Gradient" approximation in this context) that supposes the horizontal gradient of $\rho(c_p T + \Phi + E_k)$ is small, for drag-free simulations, the heat transport by the divergent circulation in the horizontal wind dominates, while the contribution of the rotational flow is also significant \citep{hammondRotationalDivergentComponents2021}. Their Figure 9 shows that the heat transport by the divergent circulation on a drag-free hot Jupiter results in a heating peak around the evening terminator and a cooling low on the dayside. Our TOA internal heat flux exhibits an anti-correlated pattern (\Cref{toaflx}e). The internal heat flux (\Cref{toaflx}f) in the strong-drag case is positive on the nightside and negative on the dayside, which is also in line with the divergent flow pattern dominating the horizontal circulation (\Cref{latlon2d}d).

By comparing the morphology of the RCB and the pattern of the internal heat flux at the TOA, one can gain further insights into how the internal heat is vertically transported out in hot Jupiter's atmosphere. If the local atmospheric column is in RCE, as in the traditional 1D framework, the outgoing internal heat flux should follow the RCB morphology (e.g., \citealt{zhangInhomogeneityEffectII2023}). However, the pattern of the drag-free internal heat flux (\Cref{toaflx}e) correlates better with the temperature and vertical wind at 0.1 bar than with the RCB pattern (\Cref{latlon2d}), implying that the local atmospheric columns should not be regarded as being in RCE. Instead, the fact that the internal heat pattern looks similar to that of the temperature and vertical wind supports the idea that the stratified atmosphere should be considered as a radiation-circulation zone, where atmospheric dynamics plays an important role in transporting energy upward. 

In the strong-drag case (\Cref{toaflx}f), the horizontal heat transport seems to be less important, so the internal heat flux has a clear day-night contrast pattern, similar to that of the temperature, vertical velocity, and RCB. But even in this case, the atmospheric columns are also not in RCE. A detailed look at the internal heat flux distribution shows that it more resembles the temperature and vertical velocity patterns, rather than the RCB pattern. For example, \Cref{latlon2d}h shows broadly flat RCB distributions on the dayside and nightside, respectively. But the internal heat flux pattern has a clear substellar-to-anti-stellar distribution (\Cref{toaflx}f). This implies that the TOA heat flux is significantly modulated by the atmospheric dynamics above the RCB, as also shown in the global-mean sensible heat flux (\Cref{eflx}).

With the same interior entropy, the normalized global-mean internal heat flux in the local case is approximately 2.3\% by design, whereas in the 3D global simulations, it is around 3.7\% for the drag-free case and 7.1\% for the strong-drag case, respectively (\Cref{eflx}). The drag-free model produces around 50\% more internal heat flux than the local model, indicating that a greater inhomogeneity in the 3D geometry would enhance internal cooling. The strong-drag case, which has a strong day-night contrast and less heat redistribution, generates more than three times the internal flux of the local model and twice that of the drag-free model. If the hot Jupiter is in energy equilibrium, much larger internal heating is needed to balance the internal cooling fluxes in 3D simulations than the value calculated in the local model.

\begin{figure*}
\centering \includegraphics[width=0.95\textwidth]{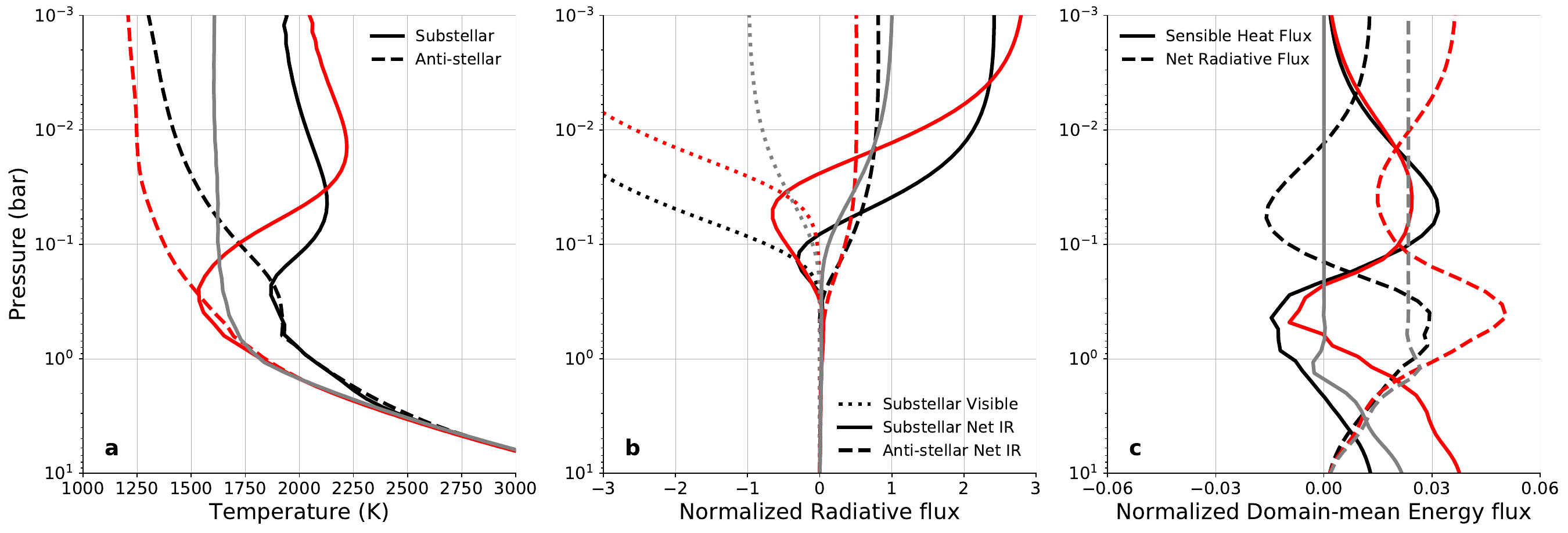}
\caption{Comparison of the local model (grey) and global models with temperature-independent opacity (black) and that with temperature-dependent opacity (red) for $T_{eq}=1600$ K. (a) Vertical temperature profiles at the substellar (solid) and anti-stellar (dashed) points. The local model has only one vertical profile. (b) Normalized radiative fluxes in the visible band at the substellar point (dotted), net IR radiative flux (upward+downward) at the substellar point (solid), and the net IR flux at the anti-stellar point (dashed). The global-mean stellar flux in the local model is 1 at the TOA. (c) Sensible heat flux (solid) and totally net radiative flux (visible + IR, dashed).}\label{Tradflx}
\end{figure*}
These results appear to align qualitatively with the proposed inhomogeneity effect and analytical prediction in \citet{zhangInhomogeneityEffectInhomogeneous2023,zhangInhomogeneityEffectII2023} using the 1D RCE framework. In 1D models, a higher RCB typically implies a larger internal heat flux, and this trend appears to hold in comparing the local and 3D global simulations. In the simulations described here, the RCB altitudes increase from the local model (3 bar) to the drag-free case (1.5 bar) to the strong-drag case (1 bar), consistent with the increasing trend of internal heat flux in the three cases. This observation is true at least for the $T_{eq}=1600$ K cases.

 In sum, we have revealed a complex picture of dynamical heat transport that is closely connected to the internal heat flux observed at TOA. The vertical heat flux is primarily transported by convection in the convective zone and regulated by radiation, large-scale circulation, and waves in the overlying stratified part. Above the RCB, several regions are identified according to their different heat transport mechanisms. The temperature inversion occurs on the dayside and inhibits radiative flux, atmospheric dynamics relay the vertical heat transport until radiation becomes efficient again. The efficiency of vertical heat transport critically depends on the correlation between the global distributions of vertical velocity and temperature patterns. Horizontal heat transport is closely related to vertical heat transport via the continuity requirement. When integrated along the atmospheric column, the convergence of the horizontal energy flux would lead to more local internal heat flux escaping from the TOA and vice versa. 
 
 However, the physical mechanisms causing the differences in interior flux among local, drag-free global, and strong-drag global models have not been fully elucidated. In addition to atmospheric dynamics, the spatial inhomogeneity of opacity plays a critical role in radiative energy transport, which we will elaborate on further in the next section.

\section{Effects of Opacity} \label{sec:opa}

As illustrated in \Cref{eflx}, the dynamically transported sensible heat flux and the radiative flux are two aspects of the same problem of interior heat transport. The latter implies that opacity plays a critical role. Due to limitations of the analytical framework, \citet{zhangInhomogeneityEffectInhomogeneous2023,zhangInhomogeneityEffectII2023} only briefly investigated the opacity inhomogeneity effect. In this section, we first simulate cases with temperature-dependent opacity and compare them to those with temperature-independent opacity. This experiment enables us to separate the roles of radiation and dynamics in heat transport and highlights the importance of opacity inhomogeneity in the atmosphere of hot Jupiters. Second, we run GCM simulations with different visible-to-IR opacity ratios and explore the internal heat flux in a larger parameter space.

\subsection{Impacts of the Temperature-dependence of Opacity} \label{sec:TDO}

The simulations with temperature-independent opacity were set up in the same manner as in the previous section, with the exception of using \Cref{TIopa} for the IR opacity, which is only pressure-dependent. The local model with the new opacity was first confirmed to reproduce the results of the local model in Section \ref{sec:local}, since the horizontal temperature variation was minimal. Then, drag-free and strong-drag global cases were simulated. The typical weather patterns and the vertical profiles of the global-mean vertical energy fluxes in both the drag-free and strong-drag cases generally resembled those in their counterparts with temperature-dependent opacity and thus not shown here.

A striking result was that for models with temperature-independent opacity, the global mean internal heat fluxes were significantly smaller---only 1.3\% for the no-drag case and 2.8\% for the strong-drag case. This is in contrast to the temperature-dependent cases, which had a much larger internal heat flux - 3.7\% for the no-drag case and 7.1\% for the strong-drag case. The flux in the drag-free case is even smaller than in the local model (2.3\%), while the flux in the strong-drag case appeared slightly larger. This implies that the larger interior cooling from the 3D global simulations in Section \ref{sec:global} might be primarily caused by the temperature-dependence nature of the opacity, and that atmospheric dynamics may act to transport internal heat flux downward.

How does the temperature dependence of IR opacity change the interior flux transport in 3D global simulations? Because the opacity directly affects radiation, we illustrated the temperature profiles at the substellar and anti-stellar points from two drag-free models in \Cref{Tradflx}a. We focus on the region from 10 bar to 1 mbar to zoom in on the ``radiation-circulation" zone. The substellar temperature profiles in both cases show clear inversion, while the anti-stellar profiles do not. But the day and night temperatures below the 0.1 bar level in the temperature-dependent opacity simulation (red) are significantly colder and decrease faster with altitude than those with temperature-independent opacity (black). Moreover, above the 20 mbar level, the dayside temperature from the case with temperature-dependent opacity is warmer than that with temperature-independent opacity, while the nightside temperature is colder. In other words, the day-night temperature contrast in the case with temperature-dependent opacity is much larger. As a result, the vertical radiative fluxes in the two simulations were significantly different (\Cref{Tradflx}b). The RCB in the case with temperature-independent opacity was much deeper than that with the temperature-dependent opacity (\Cref{Tradflx}c).
\begin{figure*}
  \centering \includegraphics[width=0.95\textwidth]{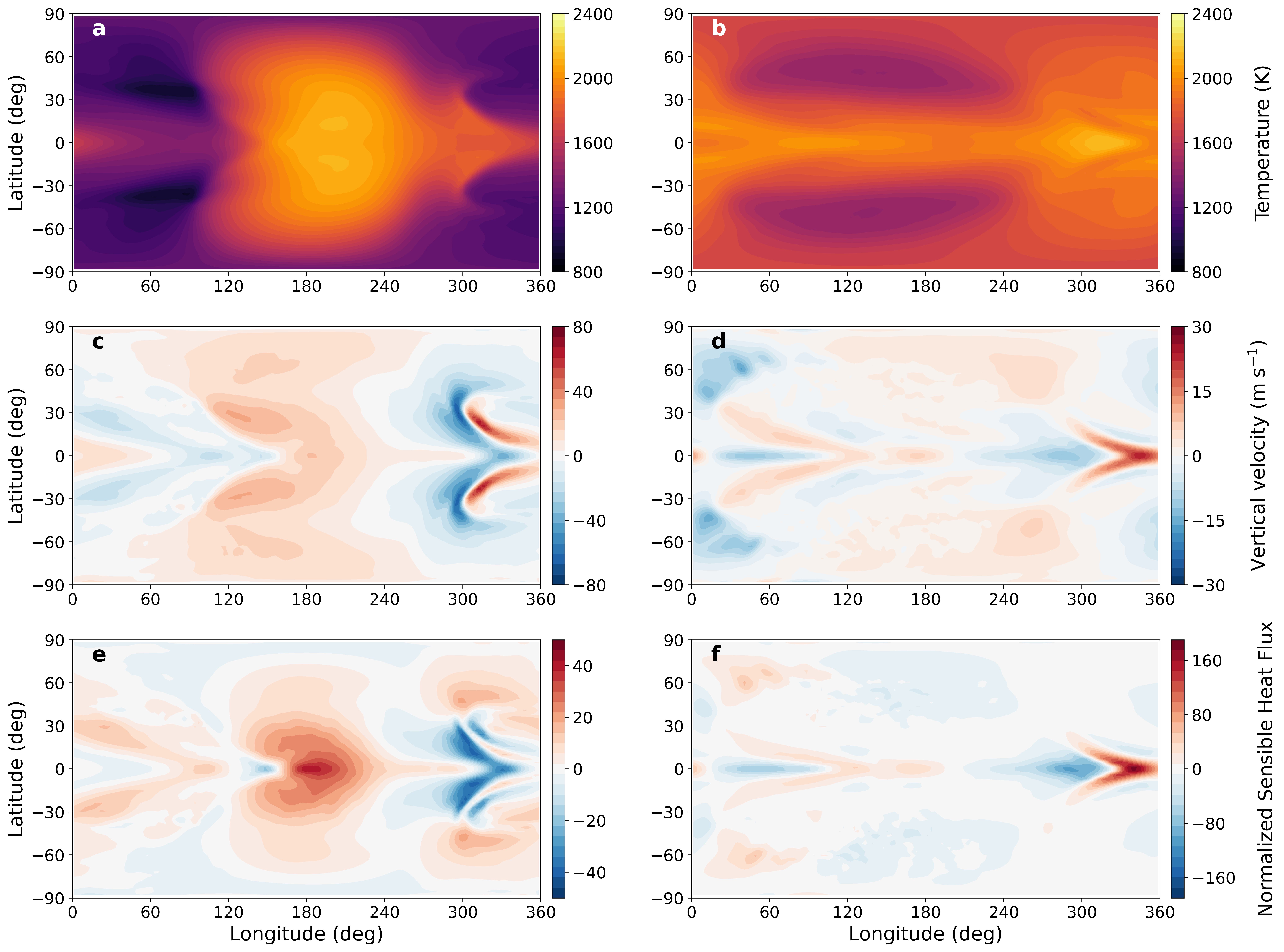} 
  \caption{Global maps of the temperature (top row), vertical velocity (middle row), and the normalized sensible heat flux (bottom row) for a 3D global simulation with $T_{eq}=1600$ K and temperature-independent opacity (Equation \ref{TIopa}). We show the maps at two pressure levels: 0.05 bar (left) and 0.5 bar (right).}\label{wtflx} 
\end{figure*}

The temperature distribution leads to a large opacity inhomogeneity and changed the vertical radiative fluxes (\Cref{Tradflx}b and c). Compared with the temperature-independent opacity that corresponded roughly to the temperature at 1600 K, the realistic IR opacity from \citet{freedmanGaseousMeanOpacities2014} significantly increases on the dayside with the thermal inversion ($>$2200 K), while the opacity significantly decreases on the nightside ($\sim$1200 K, see \Cref{opa}). As the visible opacity scaled with the IR opacity in the simulations, the dayside stellar flux is quickly absorbed in high altitudes if the opacity is temperature-dependent (\Cref{Tradflx}b, red dotted).

With the same internal entropy, a larger stellar flux absorption at higher latitudes would allow more IR emission from the deep atmosphere. But this is not the reason why the internal heat flux is larger in the simulation with temperature-dependent opacity because the IR opacity increase would play a larger role in inhibiting the flux out. As shown in 1D RCE models (\citealt{zhangInhomogeneityEffectInhomogeneous2023}), increasing \textit{both} the visible and IR opacity would actually reduce the interior cooling. Instead, the internal heat flux in the above 3D simulation with temperature-dependent opacity is caused by the spatial inhomogeneity from the dayside to the nightside, as explained below.

As the visible opacity increases on the dayside, the underlying atmosphere becomes colder with a large temperature gradient with height on both dayside and nightside (\Cref{Tradflx}a, red). A larger temperature gradient and a smaller nightside IR opacity significantly enhance the cooling of the atmosphere. The nightside IR flux exceeds that on the dayside (\Cref{Tradflx}b, red dashed). On the contrary, in the case of temperature-independent opacity, the stellar flux penetrates much deeper on the dayside and efficient day-night heat transport warms up the nightside atmosphere but reduces its vertical temperature gradient (\Cref{Tradflx}a, black). The net effect inhibits the upward transport of IR radiation in the global-mean sense (\Cref{Tradflx}b, black dashed). Furthermore, because the downward visible flux in this region is much smaller in the case of temperature-dependent opacity, the net upward radiative flux (visible+IR) becomes even larger due to less heating (\Cref{Tradflx}c). 

Above around 0.1 bar (the second region of the ``radiation-circulation zone"), the dayside thermal inversion yields downward IR fluxes in both cases (\Cref{Tradflx}b) and reduces the net upward IR radiative flux in the global-mean sense (\Cref{Tradflx}c). Due to a stronger thermal inversion, the temperature-independent opacity case has a smaller IR radiative flux above 0.1 bar, reaching the minimum at 50 mbar. However, because a larger stellar flux further intensifies the downward radiative flux in the case of temperature-independent opacity, the global-mean net radiative flux becomes negative (\Cref{Tradflx}c, black dashed). In this region, the circulation plays a crucial role in transporting the sensible heat flux outward. The upward flow carries the hotter air on the dayside, while the downward flow carries the colder air on the nightside. 

The net flux is upward by dynamics until the radiation dominates the heat transport again in the low opacity region (the third region of the ``radiation-circulation zone"), at around 10 mbar above the thermal inversion layer (\Cref{Tradflx}c). Compared with the temperature-dependent opacity simulation, the dayside IR flux in the temperature-independent opacity case eventually becomes larger above 5 mbar as the dayside temperature is higher at the top of the atmosphere, but the nightside emission flux remains lower due to its colder temperature.

The general behavior is similar in the strong-drag case. But day-night contrast is larger. Inefficient day-night heat transport in the presence of strong drag leads to a cold nightside temperature, significantly reducing the IR opacity in the temperature-dependent case and allowing most internal heat flux to be emitted outward. The above argument is consistent with the ``opacity inhomogeneity effect" proposed in \citet{zhangInhomogeneityEffectInhomogeneous2023} and explains why the strong-drag case emits more internal heat flux in the 3D global simulations than the drag-free case.

\subsection{Roles of Dynamical Heat Transport} \label{sec:dyn}

However, it is puzzling that the 3D global simulation with temperature-independent opacity transports less internal heat flux outward than the local model with the same internal entropy and atmospheric opacity. This result contradicts the inhomogeneity effect proposed in \citet{zhangInhomogeneityEffectInhomogeneous2023,zhangInhomogeneityEffectII2023}, which suggests that the spherical geometry of giant planets would lead to the inhomogeneity of the incoming stellar flux across the globe, thereby increasing the cooling of the planet's interior. It appears that radiation may not be the cause of the internal heat flux difference because the temperature-independent opacity in the local and global models follows almost the same distribution as a function of pressure. Thermal inversion is also observed in RCE models but did not cause a significant interior heating (\citealt{zhangInhomogeneityEffectII2023}). Instead, it suggests that an additional mechanism, not included in the 1D RCE framework of \citet{zhangInhomogeneityEffectInhomogeneous2023,zhangInhomogeneityEffectII2023}, is responsible for less internal heat flux difference in the drag-free 3D global simulation.

Given the downward sensible heat flux above the RCB in the simulation with temperature-independent opacity (\Cref{Tradflx}c), this missing mechanism is most likely associated with atmospheric dynamics. We examined the temperature and vertical wind distributions at around 0.05 and 0.5 bar, as well as the global maps of the local sensible heat flux (defined as $\rho c_p w T$) at these levels (\Cref{wtflx}). At the upper level, well above the RCB, the temperature pattern and vertical velocity show a good spatial correlation. The dayside and most of the nightside exhibit upward heat flux, except for some localized downwelling regions near the evening terminator at the equator. However, at a lower level around 0.5 bar, the temperature pattern is different. The temperature is warmer in the equatorial region, with two cold regions at middle and high latitudes spanning from 60 to 240 degrees in longitude (\Cref{wtflx}b). At the same level, the vertical wind shows a wavenumber-two eddy pattern with both upward and downward plumes at the equator and two large downwelling regions in the middle and high latitudes (\Cref{wtflx}d). Consequently, the upward and downward sensible fluxes at the equator almost cancel each other out (\Cref{wtflx}f). The net vertical heat transport is primarily dominated by the downward flux in the middle and high latitudes (\Cref{wtflx}f). This large (global-mean) downward heat transport, which does not exist in the local models (\Cref{Tradflx}c), is responsible for the smaller upward internal heat flux at the TOA.

In principle, the downward flux transport by atmospheric dynamics should also be present in the simulations with temperature-dependent opacity since the temperature pattern in Figure (\ref{wtflx}b) resembles the RCB distribution in Figure \ref{latlon2d}g, which traces the temperature distribution right above the convective region. The low-temperature lobes at the mid-latitudes are also seen in some simulations in recent GCM work with hot interiors (e.g., the temperature map at 1 bar in Figure 1 of \citealt{komacekEffectInteriorHeat2022}). The global-mean sensible heat flux is also slightly negative at around 0.5 bar (see the red solid line in \Cref{Tradflx}c). However, the aforementioned opacity inhomogeneity effect significantly enhanced the radiative flux and dominates the heat transport in that case, leading to a larger internal heat flux than the local model. We will delve into the deeper implications of the dynamical heat transport on the interior evolution of hot Jupiters in Section \ref{heatmechanism}.

\subsection{Dependence on the Visible-to-IR Opacity Ratio} \label{sec:alpha}
All simulations presented above assumed a unity ratio of visible to IR opacity. However, in reality, the two types of opacity might not have any significant dependence on each other. In order to test the robustness of our results, we varied the visible-to-IR opacity ratio ($\alpha$) in simulations of $T_{eq}=1600$ K. Specifically, we considered values of $\alpha=0.3$, 1, and 3.

Varying the visible-to-IR opacity ratio would result in a change in the outgoing internal heat flux in the local model. Therefore, we adjusted the bottom temperature in the local models to reproduce the same internal heat flux value (i.e., 2.3\% for $T_{eq}=1600$ K). We found that the bottom temperature for temperature-dependent opacity runs was 9800 K and 7100 K for $\alpha=0.3$ and 3, respectively. For temperature-independent opacity runs, the required bottom temperatures were slightly different, as we only adopted the fitted analytical opacity from \Cref{TIopa} which was based on the temperature profile in the case of $\alpha=1$ (i.e., we did not refit the opacity profiles for different $\alpha$). For $\alpha=0.3$ and 3, we adjusted the bottom temperature to be 8900 K and 7800 K, respectively, to achieve the same internal heat flux value as in the other cases. The required bottom temperature decreased with increasing $\alpha$ because a larger visible-to-IR opacity ratio implies a larger absorption of the stellar flux at high latitudes, resulting in a larger interior cooling. In order to yield the same internal heat flux at the TOA, the interior entropy had to be reduced.
\begin{figure}
  \centering \includegraphics[width=0.47\textwidth]{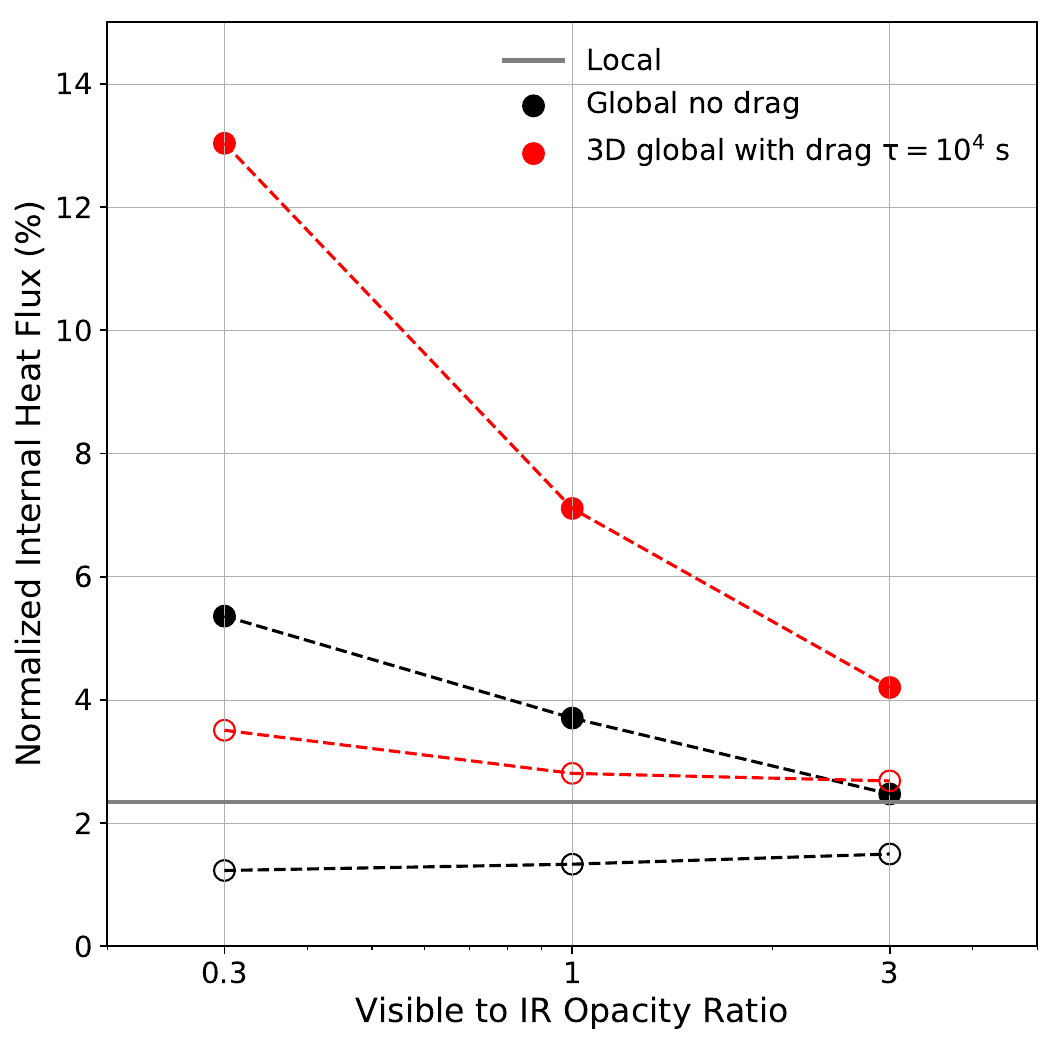} 
  \caption{Normalized internal heat flux as a function of visible-to-IR opacity ratio for drag-free (black) and strong-drag (red) simulations of $T_{eq}=1600$ K. The filled and open circles represent the simulations with temperature-dependent and temperature-independent opacity, respectively. The grey line indicates the local model with a normalized heat flux of 2.3\%.}\label{visIRratio} 
\end{figure}

We ran two global simulations for each $\alpha$ value and for both temperature-dependent and temperature-independent opacity, with and without strong drag. The normalized mean internal heat fluxes at TOA for these $3 \times 4$ cases are shown in \Cref{visIRratio}. As expected, the simulations with temperature-dependent opacity yielded much larger internal heat flux values than those with temperature-independent opacity in each drag category, due to the opacity inhomogeneity effect. The strong-drag cases produced more significant internal heat fluxes than the drag-free cases for each opacity category, as the inhomogeneity effect increased. All the drag-free global simulations with temperature-independent opacity yielded internal heat flux values that were roughly the same and lower than the results obtained in the local models, indicating a significant and robust effect of the downward heat transport by atmospheric dynamics above the RCB (Section \ref{sec:dyn}).

The internal heat flux generally decreased with increasing visible-to-IR opacity ratio. This is mainly because the interior entropy is higher for small $\alpha$, as increasing visible opacity would generally reduce the interior cooling. Among the cases considered, the largest internal heat flux is produced when the drag is strong and $\alpha=0.3$. The normalized flux value reaches about 13\%, about six times that obtained in the local models. This value of 13\% suggests that the internal temperature could be about 60\% of the equilibrium temperature, even though in the local model the internal temperature could be incorrectly estimated as only 40\% of $T_{eq}$ (corresponding to 2.3\%). We expect that the internal heat flux could further increase if we decreased $\alpha$, causing even larger differences in the interior cooling between the local and global models.

\section{Implications on Observations and Interior Heating Mechanisms} \label{sec:implication}

Correctly estimating the internal heat flux of hot Jupiters requires characterizing the spatial distribution of their thermal emission, which exhibits large spatial variation. Both the dayside low and nightside high are important for this purpose because the residues are what matter (\Cref{toaflx}e and f). A possible approach to mapping the spatial distribution of thermal emission is through phase curve measurements covering enough wavelengths to encompass the infrared emission from the planet. However, the small fraction of the internal heat flux relative to the incoming stellar flux might make it difficult to directly extract the difference between the infrared emission and the incoming stellar flux. Moreover, the planetary bond albedo is a large source of uncertainty, and previous studies have usually neglected the internal heat flux and used the day-night contrast to estimate the bond albedo \citep[e.g.,][]{cowanStatisticsAlbedoHeat2011,keatingUniformlyHotNightside2019}.

\subsection{Drag Effect on the Interior Cooling}

If it is challenging to constrain the internal heat flux for individual planets, a statistical approach might be more practical, like the work in \citet{thorngrenBayesianAnalysisHotJupiter2018} and \citet{sarkisEvidenceThreeMechanisms2021}. However, their 1D approach neglects atmospheric inhomogeneity on the interior cooling and assumes full heat distribution. Our work suggests that one can improve the method by including the spatial variability, an observable that was not used in the 1D framework but can be constrained from phase curve observations (day-night contrast, (\citet{parmentierExoplanetPhaseCurves2018}), limb transmission observations (morning-evening contrast, \citet{powellTransitSignaturesInhomogeneous2019}), and more detailed eclipse mapping (e.g., \citealt{williamsResolvingSurfacesExtrasolar2006,rauscherEclipseMappingHot2007}).

\begin{figure}
  \centering \includegraphics[width=0.47\textwidth]{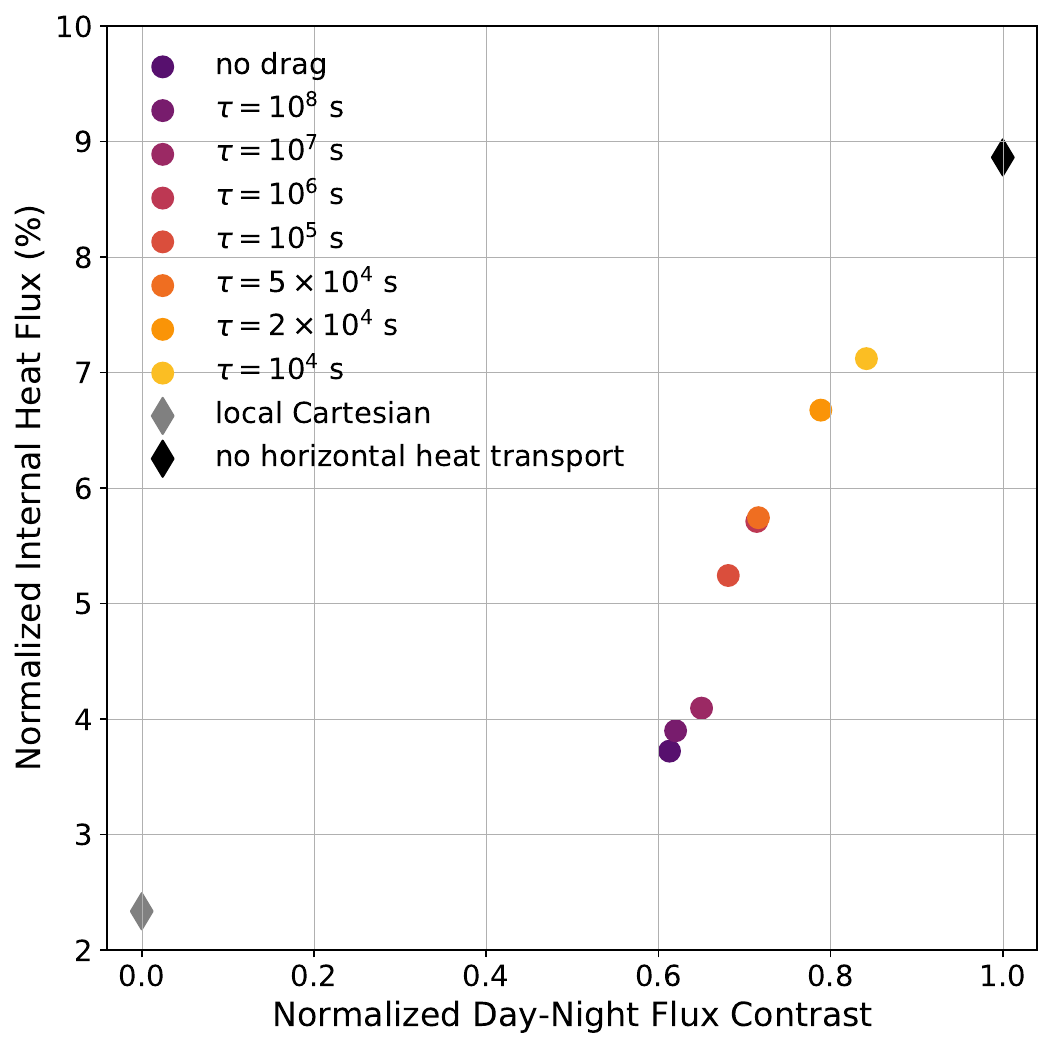} 
  \caption{Normalized internal heat flux as a function of day-night Flux contrast (Equation \ref{eqflxdiff}). The drag timescales used in the global simulations of $T_{eq}=1600$ K are labeled with different colors. The local Cartesian model (grey diamond) is located at the lower left with zero contrast and a normalized heat flux of 2.3\%. We also showed a 3D model result (black diamond) without horizontal heat transport (i.e., every local column is in RCE), in which the normalized heat flux is about 8.9\%.}\label{daynightdiff} 
\end{figure}

The observed spatial inhomogeneity on hot Jupiters can be leveraged to improve the estimate of the statistical distribution of the heating efficiency, given the relationship between the internal heat flux and inhomogeneity in 3D global simulations. However, the origin of spatial inhomogeneity on hot Jupiters is quite complex. For instance, the thermal phase curves are influenced by a variety of atmospheric processes, such as atmospheric dynamics, radiation, chemistry, and cloud formation (e.g., \citealt{komacekAtmosphericCirculationHot2016,komacekAtmosphericCirculationHot2017,parmentierExoplanetPhaseCurves2018,zhangAtmosphericRegimesTrends2020}, and particularly a comprehensive recent study in \citealt{parmentierCloudyShapeHot2021}).

As an initial exploration, in our cloud-free, grey GCM framework, we ran a series of simulations with varying atmospheric drags in order to directly modify the day-night contrast and heat redistribution. We used the same setup as in our nominal $T_{eq}=1600$ K cases with a visible-to-IR opacity ratio of unity and varied the drag timescale to $10^4$ s (the nominal ``strong drag" case), $2\times 10^4$ s, $5\times 10^4$ s, $10^5$ s, $10^6$ s, $10^7$ s, $10^8$ s, and $\infty$ (the nominal ``drag-free" case). For each simulation, we calculated the disk-averaged IR emission flux assuming the sub-observer point is at the substellar point for the dayside flux ($F_d$) and at the anti-stellar point for nightside flux ($F_n$), respectively. We then defined the day-night flux contrast ($\zeta$) as follows:
\begin{equation}
\zeta = \frac{F_d-F_n}{F_d+F_n}.
\label{eqflxdiff}
\end{equation}
This definition considers an atmosphere with full heat redistribution ($F_d=F_n$) as having $\zeta=0$, and an atmosphere with no heat redistribution ($F_n$=0) as having $\zeta=1$. A larger $\zeta$ indicates a larger day-night contrast and larger spatial inhomogeneity in the atmosphere. 

Our approach to heat redistribution efficiency, $\zeta$, differs from earlier definitions that are based on the assumption of radiative equilibrium. In those definitions, such as in \citet{spiegelATMOSPHERESPECTRALMODELS2010} and \citet{cowanStatisticsAlbedoHeat2011}, total emitting flux remains constant, irrespective of changes in heat redistribution. However, in our GCM simulations, the internal heat flux varies with the degree of atmospheric inhomogeneity, which means that earlier definitions of heat redistribution efficiency do not strictly apply. Moreover, $\zeta$ is related but does not directly correspond to the amplitude of the bolometric phase curve because the maximum and minimum points on the phase curve could be Doppler shifted by the zonal jets and modulated by waves on hot Jupiters, resulting in offsets from the substellar and anti-stellar points, respectively (e.g., \citealt{zhangEffectsBulkComposition2017,hammondWavemeanFlowInteractions2018,wangPhaseShiftPlanetary2021}).

As the atmospheric drag increases, the day-night heat redistribution generally decreases, leading to a larger day-night contrast and larger interior cooling (\Cref{daynightdiff}). This result further confirms the inhomogeneity effect in the analytical framework in \citet{zhangInhomogeneityEffectII2023}. It is also consistent with the previous 1+1D (dayside+nightside with some heat redistribution) hot Jupiter evolution models, which found that less heat redistribution would increase the interior cooling (e.g., \citealt{guillotEvolution51Pegasus2002,budajDayNightSide2012,spiegelThermalProcessesGoverning2013}). The day-night flux contrast $\zeta$ ranges from 0.61 in the drag-free case to 0.84 in the strong-drag case with a drag timescale of $10^4$ s. The range is much narrower than the theoretical limit of 0-1. Apparently, even without any drag, the atmosphere could not fully homogenize the temperature distribution across the longitude because the radiative timescale is not long enough to be neglected compared with the dynamical heat transport, as shown in the temperature pattern in \Cref{latlon2d}a. The normalized internal heat flux in the drag-free case ($\zeta=0.61$) is about 3.7\%. In the strong-drag case with a drag timescale of $10^4$ s, the heat redistribution by the zonal dynamics is greatly reduced. However, the divergent flow could still efficiently transport heat from the dayside to the nightside via the substellar-to-anti-stellar circulation, as depicted in (\Cref{latlon2d}d). In this strong-drag case ($\zeta=0.84$), the normalized internal heat flux is much larger, reaching above 7\%. 

For comparison, we plotted two extreme scenarios. In a ``static" limit scenario, strong atmospheric drag suppresses all horizontal motion above the RCB. This leads to zero horizontal heat redistribution ($\zeta=0$), essentially turning all atmospheric columns on the 3D globe into independent RCE columns. By using the local Cartesian model, we can estimate the internal heat flux for each column with a day-night stellar flux distribution. The normalized internal heat flux for this 3D ``static" model is around 8.9\%. This aligns with the trend we see in the 3D global simulations shown in \Cref{daynightdiff}, which can be extrapolated to reach the ``static" limit.

On the other extreme, the 2D/3D local model assumes full heat redistribution across the globe ($\zeta=1$), which results in an internal heat flux of 2.3\%. However, the trend displayed by the 3D global simulations in \Cref{daynightdiff} does not align with this full-redistribution value. In other words, the local model does not seem to represent the theoretical limit of the 3D global model even if we were able to entirely homogenize the longitudinal temperature contrast on tidally locked exoplanets. This discrepancy arises because atmospheric radiation becomes efficient enough to transport heat and control temperature in the upper atmosphere preventing the atmosphere on 3D tidally locked planets from reaching a state of full heat redistribution. Whether a fast-rotating planet would reach the local model limit is an interesting question and deserves a separate study in the future.

\subsection{Revisit the Temperature Dependence of Heating Efficiency}

The opacity inhomogeneity effect depends critically on the functional shape of opacity with temperature (\Cref{opa}), and the dependence of interior flux on drag is expected to differ for different equilibrium temperatures. Therefore, to investigate the weather impacts across a larger parameter space, we performed simulations for planets with $T_{eq}$ ranging from 1200 K to 2200 K in 200 K increments, assuming a visible-to-IR opacity ratio of unity. We followed the same approach as in our nominal cases with $T_{eq}=1600$ K. First, we found the bottom temperature in our local models to reproduce the previously derived heating efficiency for each $T_{eq}$. Then, we used the same bottom temperatures (as a proxy for interior entropy) in global models and simulated both drag-free and strong-drag cases, as in Section \ref{sec:global}.

The derived heating efficiencies in \citet{thorngrenBayesianAnalysisHotJupiter2018} and \citet{sarkisEvidenceThreeMechanisms2021} are roughly comparable below 1700 K, above which the efficiency is larger than the former (see Figure 6 in \citealt{sarkisEvidenceThreeMechanisms2021} and \Cref{teqdep}). To simplify the comparison, we chose to compare with \citet{thorngrenBayesianAnalysisHotJupiter2018} in this study because the difference between our local and global models could be much larger than their 1D model difference, as shown in our $T_{eq}=1600$ K cases. The derived bottom temperatures are listed in Table \ref{tablepara}, and the global-mean normalized internal heat fluxes are summarized in \Cref{teqdep}.

The 3D simulations reveal interesting differences when compared to the local models. Specifically, in the drag-free models, only the $T_{eq}=1600$ K and 1800 K cases emit more internal heat flux than the local models. The fluxes from the hotter models ($T_{eq} > 1800$ K) are comparable to the local models, while the fluxes from the colder models ($T_{eq} < 1600$ K) are slightly smaller due to the downward heat transport by atmospheric dynamics. For the $T_{eq}=1600$ K and 1800 K cases, the effect of temperature-dependent opacity plays a larger role and increases the internal heat flux. This is supported by the grey IR opacity plot in \Cref{opa}, which shows that the $T_{eq}$ range from 1600 to 1800 K is situated between the low opacity region in the colder regime and the high opacity region in the hotter regime. Thus, in these two simulations, the opacity on the hotter dayside is substantially larger than that on the colder nightside. This leads to the inhibition of the downward visible flux on the dayside and the enhancement of the upward IR flux on the nightside, resulting in a substantial boost to the interior cooling. Conversely, if the equilibrium temperature is colder, such as $T_{eq}=1200$ K, the dayside and nightside may have roughly the same opacity due to the concave shape of the opacity contour. Similarly, in hotter equilibrium temperature regimes such as 2000 K, the opacities on the day and nightsides are comparable, and the opacity inhomogeneity effect is not as pronounced.

\begin{figure}
  \centering \includegraphics[width=0.47\textwidth]{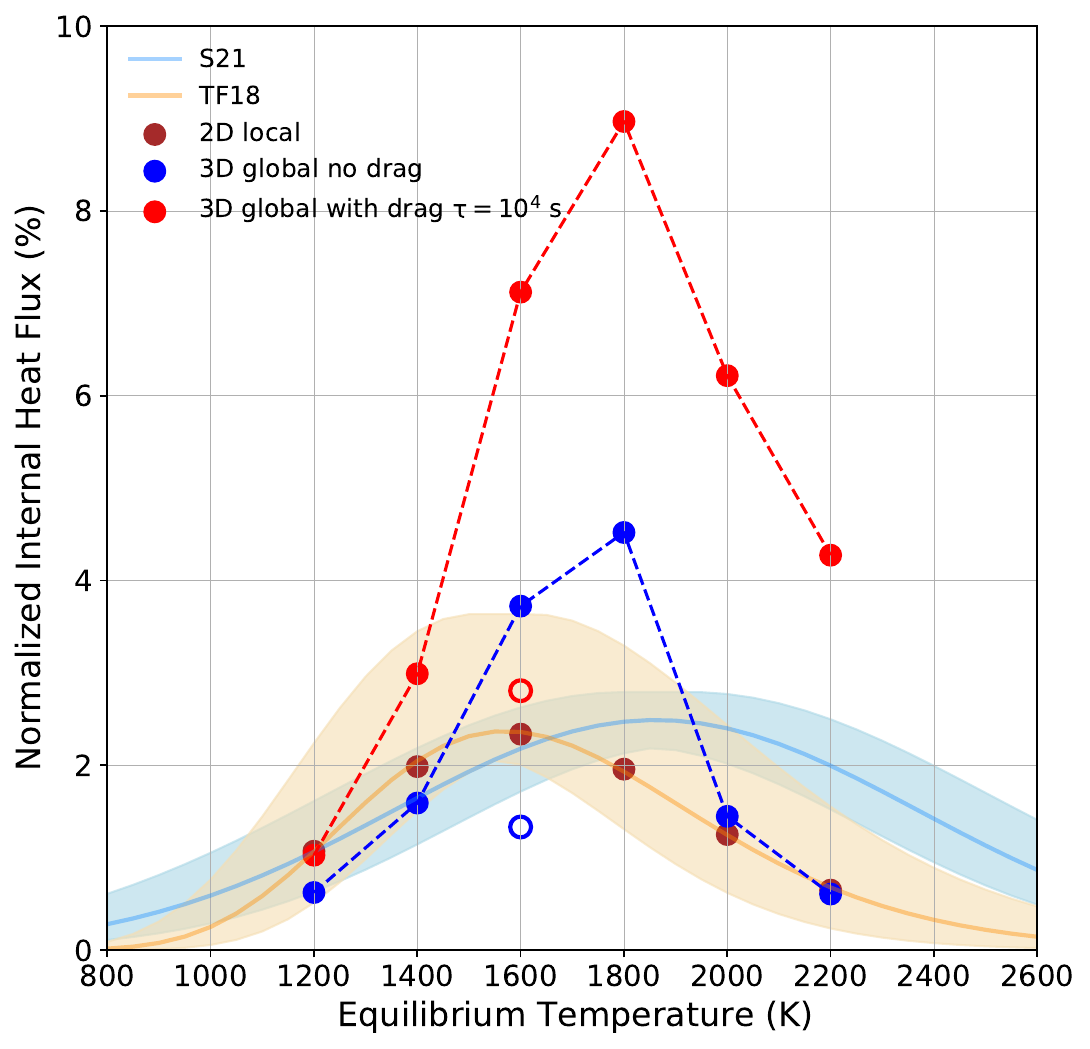} 
  \caption{Normalized internal heat flux (i.e., the required heating efficiency to inflate the hot Jupiter) as a function of equilibrium temperature from 2D local (brown), 3D global drag-free (blue), and strong-drag (red) models with temperature-dependent opacity and visible-to-IR opacity ratio of unity. The two open circles represent the heat fluxes from simulations with temperature-independent opacity and visible-to-IR opacity ratio of unity. We also show heating efficiencies and shaded uncertainty ranges from previous statistical analysis based on 1D models in \citet{thorngrenBayesianAnalysisHotJupiter2018} (``TF18", orange) and \citet{sarkisEvidenceThreeMechanisms2021} (``S21", light blue). The bottom temperatures in the local models are tuned to reproduce the efficiencies from \citet{thorngrenBayesianAnalysisHotJupiter2018}. }\label{teqdep} 
\end{figure}

If the drag is strong, spatial inhomogeneity increases and the internal heat flux increases in all cases. However, this effect is not very significant for colder planets, such as the one at 1200 K, where the strong-drag case yields an internal heat flux similar to that in the local models. For 1400 K planets, the internal heat flux is about 50\% larger than that in the local model. But for hotter planets beyond 1600 K, the drag effect is significant. For the 2200 K case, the strong-drag simulations produce an internal heat flux of about 4.2\%, seven times larger than the local/1D model results of 0.6\% in \citet{thorngrenBayesianAnalysisHotJupiter2018}.

If the planets at all temperature share the atmospheric drag, the overall shape of the normalized internal heat flux as a function of temperature from the global models is similar to the curve from the local model (\Cref{daynightdiff}). The curves show a peak in the middle temperature and a decrease towards the hot and cold end. This is probably driven by the bottom temperature specified in these cases to reproduce the 1D model heating efficiency in the local models. The flux peaks at $T_{eq}=1800$ K in the 3D global cases, slightly different from that in the local model ($T_{eq}\sim1600$ K, \citealt{thorngrenBayesianAnalysisHotJupiter2018}). 

Assuming the hot Jupiters are in energy equilibrium, the required heating efficiency is the same as the normalized internal heat flux. If we just consider the temperature dependence of the efficiency, a peak at $T_{eq}=1800$ K might be consistent with three proposed mechanisms of hot Jupiter inflation according to \citet{sarkisEvidenceThreeMechanisms2021}---Ohmic heating, downward advection of entropy, and thermal tides. For instance, the Ohmic heating efficiency increases with the ionization in the cold regime but decreases as the drag becomes stronger in the hot regime. Thus, the expected heating peak is located around $T_{eq}\sim1500$ K (\citealt{batyginEvolutionOhmicallyHeated2011,menouMagneticScalingLaws2012,rogersMagneticEffectsHot2014,ginzburgExtendedHeatDeposition2016}), a temperature slightly lower than the peaks in our global simulations with/without the same drag.

However, in real atmospheres, the drag is probably not constant in the entire atmosphere, as we have assumed here (\citealt{beltzExploringEffectsActive2022}). The drag should also change with the equilibrium temperature. If it mainly comes from the magnetic effect, the atmospheric drag could be more important for a hotter atmosphere, as the ionization fraction is higher. Therefore, the local model might have slightly overestimated the heating efficiency for cold planets due to the lack of dynamical heat transport but largely underestimated the efficiency for hot planets due to the lack of inhomogeneity effect (\Cref{daynightdiff}). If we take this effect into account, the peak of the heating efficiency curve might be considerably shifted towards high temperature or even disappear.

The effect of the visible-to-IR opacity ratio can further complicate the picture (\Cref{visIRratio}). In extreme cases, the strong-drag case can produce a large internal heat flux exceeding 10\%, such as in the low visible-to-IR opacity ratio case in \Cref{visIRratio}. This value is much larger than the previous estimate and could challenge the inflation mechanisms. Moreover, this efficiency is derived under the assumption that all the unknown heating is deposited in the deep interior. If the heating depth is shallower, the required heating efficiency should be even larger (e.g., \citealt{komacekStructureEvolutionInternally2017}). However, since our simulations are idealized with grey opacity and fixed visible-to-IR opacity ratios, we do not want to overinterpret our findings here.

\subsection{Implications on Inflation Mechanisms via Hydrodynamics} \label{heatmechanism}

Our non-hydrostatic atmospheric models provide a promising avenue for investigating some potential size inflation mechanisms in Hot Jupiters, which posit that atmospheric energy is transported downward to the deep interior via hydrodynamics. Such mechanisms include thermal tides \citep{arrasThermalTidesFluid2010,guModelingThermalBulge2019}, mechanical greenhouse (\citealt{showmanAtmosphericCirculationTides2002, guillotEvolution51Pegasus2002, youdinMechanicalGreenhouseBurial2010}), and heat transport by general circulation and waves (\citealt{showmanAtmosphericCirculationTides2002, guillotEvolution51Pegasus2002, tremblinAdvectionPotentialTemperature2017,sainsbury-martinezIdealisedSimulationsDeep2019,mendonca2020,schneiderNoEvidenceRadius2022,sainsbury-martinezEvidenceRadiusInflation2023}). Previous studies have mostly employed 3D GCMs to explore the heat advection hypothesis, leaving the other two hypotheses relatively unexamined.

Assuming thermal equilibrium and cessation of contraction, successful capture of these heating mechanisms by our 3D models would reflect a zero internal heat flux at the TOA. Yet, our simulations persistently demonstrate a non-zero net cooling flux, especially with temperature-dependent opacities, hinting that the hydrodynamical effects are not sufficient to offset the planetary cooling towards space in our model setup.

Our current models do not permit direct testing of the thermal tides hypothesis in the situation of asynchronous rotation and non-zero orbital eccentricity. Nevertheless, the models provide crucial insights into the viability of the turbulent kinetic energy transport and heat advection mechanisms. Our simulations do not appear to support these theories as the primary size inflation mechanisms at the resolution of this study with the global circulation pattern resolved and with the non-hydrostatic effect turned on. Although there is a possibility that smaller, unresolved eddies missed in this study might contribute to the mechanical heating of the interior, we did not observe any trends in support of this idea across the low-, mid-, and high-resolution runs conducted in this study. In addition, given the previously asserted robustness of large downward heat advection in other GCMs, our contrasting findings merit further discussion here.

\cite{showmanAtmosphericCirculationTides2002} first proposed that an equatorial circulation pattern, which rises from the dayside and sinks on the nightside, could transport the energy downward in the radiative zone. \cite{tremblinAdvectionPotentialTemperature2017} further developed the idea in their deep 2D models and proposed that the downward advective enthalpy flux was enough to explain the size inflation of hot Jupiters. This concept was examined in more detail by \cite{sainsbury-martinezIdealisedSimulationsDeep2019} using a 3D hydrostatic GCM solving primitive equations. With a Newtonian cooling scheme and an initially cold temperature profile in the lower atmosphere, they found that the lower atmosphere could be heated in their model over time. However, their focus was more on the meridional circulation that transports the entropy inward at the equator, which is different from the zonal circulation proposed by \cite{tremblinAdvectionPotentialTemperature2017}. Similarly, the non-hydrostatic simulations with temperature-independent opacities by \citet{mendonca2020} also found that the heat flux can be transported downward, resulting in a lower atmosphere about 10 K warmer than that of radiative equilibrium (see their Figure 8d).

Recently, \cite{schneiderNoEvidenceRadius2022} used a more realistic radiative transfer scheme in MITgcm and found no evidence of potential temperature being transported vertically by atmospheric dynamics, even after running the model for an extended time of 86,000 days. However, this finding was contested by \cite{sainsbury-martinezEvidenceRadiusInflation2023}, who ran the simulations for even longer and found that enough enthalpy flux, which is the same as the sensible heat flux in this study, could be carried downward to heat the interior.

Our own 3D simulations do not show the large-scale zonal circulation in the convective zone suggested by \cite{tremblinAdvectionPotentialTemperature2017}. Instead, our model shows patterns above the RCB that resemble those found by \cite{sainsbury-martinezIdealisedSimulationsDeep2019}, which demonstrate two large upward movements at high latitudes that turn into strong downward flows at lower latitudes (see their Figure 5). \citet{mendonca2020} showed similar heat flux patterns and analyzed the separate roles of the heat transport by mean flow and stationary waves (see their Figure 7). In our simulations, this circulation pattern is the main driver of the net downward heat transport above the RCB. However, the global-mean downward flow of sensible heat in our simulations is relatively small and constrained to a narrow area above the RCB. Our temperature-independent opacity simulations align with \citet{mendonca2020} findings, which indicate that the lower atmosphere's temperature exceeds that of radiative equilibrium but the temperature increase is insufficient to explain the size inflation of hot Jupiters. Moreover, the downward dynamical heat flow in our simulations is balanced out by the large upward flow of radiative heat at the same level, resulting in a constant net upward flow of energy. On the other hand, \cite{sainsbury-martinezEvidenceRadiusInflation2023} found a large peak of the downward flow of heat in the convective region in their study (see their Figure 3).

The differences between our simulation results and previous models imply intriguing areas for future exploration. First, our models solve non-hydrostatic equations with total energy advection, while the MITgcm solves primitive equations with potential temperature advection. Additionally, we have assumed grey opacity, while their model adopts a non-grey radiative scheme. Furthermore, our model setups differ significantly. We started with a fixed bottom temperature with a bottom drag and evolved the model towards a steady state of energy fluxes at the TOA, whereas the other studies started with a colder initial temperature and bottom drag, gradually evolving towards a steady-state temperature profile in the convective region.

In our simulations, the steady-state is evident from the constant vertical line for total energy flux as a function of altitude (see \Cref{eflx}), implying the energy tendency term (the time derivative of energy in Equation \ref{eneeqn}) is nearly zero. By contrast, the total energy flux in the study by \cite{sainsbury-martinezEvidenceRadiusInflation2023} appears not vertically constant across the simulation domain (see their Figure 3). This discrepancy might hint at a simulation yet to reach its steady state. Alternatively, if their fluxes were averaged over a long period after reaching the steady state, the variability in total flux with altitude might imply the need for more careful treatment in energy conservation or flux analysis. As the required heating flux to inflate the interior is only 1-10\% of the incoming stellar flux, careful spatial and temporal averaging is critical for analyzing the global-mean heat flow in these simulations. Future comparisons using the same model setup will be essential for clarifying these discrepancies and further illuminating the heating mechanisms of hot Jupiters.

In summary, our 3D global simulations reveal how weather effects can greatly alter both the magnitude and distribution of interior cooling, leading to significant deviations from the results of 1D/local models. If opacity heavily depends on temperature (as in the case of TiO/VO), a strong inhomogeneity effect can overpower downward dynamical heat transport, resulting in a higher rate of interior cooling compared to 1D models. However, when opacity has a weak dependence on temperature, the inhomogeneity effect in a 3D global model would be less pronounced than in a 1D model. Given the uncertainties, we are currently unable to definitively determine how the required heating efficiency (and intrinsic temperature) of hot Jupiters depends on temperature, which impedes our comprehension of their inflation mechanisms. While our study shows that dynamical circulation could move heat into a hot Jupiter's interior, we did not come across any scenarios where the downward heat flux completely counteracts interior cooling (i.e., our simulations did not yield negative or zero $\eta$). Therefore, our results do not support the hypothesis of potential temperature advection as the primary inflation mechanism, as suggested in previous works (\citealt{tremblinAdvectionPotentialTemperature2017, sainsbury-martinezIdealisedSimulationsDeep2019,sainsbury-martinezEvidenceRadiusInflation2023}), even for cases with temperature-independent opacity.

\section{Conclusion and Discussion} \label{sec:sum}

In this study, we utilized a 3D non-hydrostatic GCM to systematically investigate the heat flow of hot Jupiters with a double-grey radiative transfer scheme. We set the bottom temperature to approximate a homogeneous interior entropy distribution in the deep convective atmosphere and traced the energy flow from the convective interior to space in the model that directly solves the energy equation using a finite-volume scheme. Our analysis revealed substantial differences in the TOA internal heat fluxes between the local model and global models and identified important processes that shape the vertical transport of the internal heat flux. We summarize the main take-away points below:

1. The atmosphere from the 2D and 3D local models is approximately in RCE. The sensible heat flux is transported by convection in the lower atmosphere and by radiation above the RCB. The local non-hydrostatic models could be a useful tool for future RCE simulation as they are computationally efficient, and no convective adjustment is needed.

2. The global-mean heat fluxes in 3D global simulations show complex profiles. The vertical flux of kinetic energy is negligible compared to the sensible heat and radiative fluxes, which intersect multiple times from the bottom to the top of the atmosphere. Their sum is vertically constant, demonstrating good energy conservation in the model. The internal heat is transported by convection in the lower atmosphere until the RCB.

3. The RCB surface morphology is significantly impacted by atmospheric dynamics and temperature structure. If the atmospheric drag is weak, the RCB is located at a lower altitude at both the equator and high latitudes but is higher at mid-latitudes. The RCB in the evening hemisphere is also higher than in the morning. The RCB is roughly homogeneous along the longitudes at the equator due to efficient heat transport by the superrotating jet and waves. If the drag is strong, the RCB exhibits a day-night contrast pattern with low altitudes on the dayside and high altitudes on the nightside.

4. The atmosphere above the RCB in 3D global simulations is not in RCE. Both radiation and dynamics play crucial roles in transporting heat horizontally and vertically. Therefore, we term it the ``radiation-circulation" zone. This zone can be split into three regions from the bottom to the top. In the first region above the RCB, the global-mean sensible heat flux quickly decreases because temperature and vertical velocity patterns show anti-correlated patterns in the middle and high latitudes. Upward radiative flux in the infrared becomes dominant as the stellar absorption cannot penetrate deep into the dayside atmosphere, and the nightside IR emission quickly escapes to the upper atmosphere.

5. In the middle region, the dayside temperature profiles exhibit significant thermal inversion, and IR flux is radiated downward on the dayside. Together with the downward visible flux on the dayside, the downward radiative flux significantly cancels the upward flux from the nightside. On the other hand, the temperature and vertical velocity patterns show a good correlation. The dayside upwelling of hot air and the nightside downwelling of cold air efficiently transport sensible heat upward in this region.

6. In the top region, the optical depth is low, and the IR radiation becomes dominant again and is directly emitted to space on both the dayside and nightside. The dynamical heat flux quickly drops to zero as the atmospheric density decreases exponentially with altitude.

7. At TOA, the difference between the incoming stellar flux and the outgoing IR radiation shows a complex pattern of local internal heat flux, with downward flux on the dayside and outward flux on the nightside. For drag-free simulations, the evening hemisphere exhibits a larger outgoing flow. The TOA internal flux pattern follows vertically integrated horizontal energy divergence, in which the divergent flow pattern appears to dominate the horizontal heat transport (\citealt{hammondRotationalDivergentComponents2021}) and shapes the flux structure at the TOA.

8. The global-mean internal heat flux in the 3D global simulations for a $T_{eq}=1600$ K is significantly larger than in the local model. This is a result of the opacity inhomogeneity effect. In 3D global models with a large day-night temperature contrast, the IR opacity is significantly larger than the mean opacity in the local model, causing less penetration of the stellar flux on the dayside. More importantly, the nightside opacity is significantly lower than in the local model, allowing the internal heat flux to escape from the nightside like a ``radiator fin".

9. The opacity inhomogeneity effect is weak if the atmospheric opacity is temperature-independent or the day-night opacity difference is not large. In this situation, the internal heat flux in the global simulations appears close to or smaller than that in the local model because the atmospheric dynamics above the RCB transport the hot air downward in the middle latitudes, while the upward and downward vertical fluxes roughly cancel out at the equator. 

10. In all 3D global simulations, increasing the drag reduces the heat redistribution between the dayside and nightside and increases the spatial inhomogeneity. Consequently, the outgoing internal heat flux considerably increases, especially on the nightside. In the case of $T_{eq}=1600$ K, the internal heat flux scales well with the day-night IR flux contrast, suggesting the possibility of using the observed spatial variability of hot Jupiters to estimate the internal heat flux and required heating efficiency beyond the traditional 1D framework.

11. The ratio of the visible-to-IR opacity significantly impacts the internal heat flux. The flux increases as the visible-to-IR opacity ratio decreases. The internal heat flux can reach about 13\% of the absorbed stellar flux for the $T_{eq}=1600$ case with a ratio of 0.3. This is about six times that estimated in the local model. If this is accurate, we might need to combine several heating mechanisms to achieve this unusually high heating. However, note that our simulations are idealized, and the inhomogeneity effect is closely tied to atmospheric opacity. Future simulations with a non-grey opacity, which represents a more realistic scenario, should be conducted to better determine these values.

12. The required inflation heating efficiency strongly depends on the equilibrium temperature. With the same drag, the 3D global simulations show a large peak of the efficiency at around $1800$ K instead of $1600$ K in the local models if we calibrated our local models based on \citet{thorngrenBayesianAnalysisHotJupiter2018}. The local model slightly overestimates the heating efficiency for cold planets due to the lack of dynamical heat transport but largely underestimates the efficiency for hot planets due to the lack of inhomogeneity effect (\Cref{daynightdiff}). Given the significant dependence of the internal heat flux on the atmospheric inhomogeneity in the atmosphere, the previously calculated heating efficiency distribution using 1D models should be revisited.

13. In our simulations, we did not find any scenarios where the downward heat flux completely offsets interior cooling, suggesting that an additional heating mechanism other than atmospheric dynamics might be still needed to inflate the planets. Our results do not support the hypothesis of potential temperature advection by general circulation as the primary inflation mechanism in \cite{tremblinAdvectionPotentialTemperature2017}. In the future, our framework could also be used to test the thermal tides hypothesis, which involves asynchronous rotation and non-zero orbital eccentricity, and the mechanical greenhouse effect, which might require finer-resolution simulations or subgrid eddy parameterization.

We consider this work as an extension of the analytical studies conducted in \citet{zhangInhomogeneityEffectInhomogeneous2023,zhangInhomogeneityEffectII2023} with more realistic simulations. The opacity inhomogeneity effect observed in the GCMs in this study aligns with the analytical study. However, the downward heat transport by dynamics was not expected. The three papers collectively outline a general principle of planetary cooling, where planetary inhomogeneity significantly impacts the cooling of the planetary surface and interior and tends to accelerate it if the inhomogeneity is high. In the context of giant planets, atmospheric inhomogeneities will lead to a larger interior cooling. The spatial inhomogeneity is expected on tidally locked gas giants, and the dynamical circulation and temperature dependence of opacity could change the interior cooling substantially.

The simulations presented in this study are still idealized. We have introduced a technique to modify the plane-parallel radiative transfer solver to ensure flux power conservation in a spherical, extended atmosphere. In the visible band, the scheme could be justified because otherwise the energy power is not conserved with altitude. Moreover, we made this choice so that the global-mean flux in the global models and the flux in the local model have the same attenuation profile. Limb correction for high incident angles should be included in more realistic studies in the future.

In the IR band, our treatment only arguably makes sense because we did not carry out a realistic 3D spherical radiative transfer calculation. We have also performed additional simulations of $T_{eq}=1600$ K with the traditional cooling rate from the plane-parallel solver in the energy equation. We then calculated the outgoing IR flux at the TOA assuming the interior surface area is the same as that at the top level of the model, like all current GCMs do. As expected, the models yield larger cooling fluxes than the nominal models as the area in the interior is enlarged. Therefore, in this setup, the global simulations in both drag-free and strong-drag cases show even greater internal heat fluxes than the local model. The conclusion of this study would not be altered. But a more realistic, spherical radiative scheme is needed to quantitatively estimate the emission flux of hot Jupiters, especially the very hot ones with extended atmospheres.

In this study, we made certain assumptions about the opacity parameters, namely a grey IR opacity and a constant visible-to-IR opacity ratio. Given the significant impact of the opacity effect on internal cooling, any future simulations that take non-grey opacity into account should reevaluate the findings of this study. The absence of TiO/VO, one of the potential opacity sources at around 2000 K, might curtail the inhomogeneity effect due to its diminished opacity-temperature dependency. Clouds, on the other hand, introduce complexities, not only through their scattering effect but also through their multifaceted formation processes and distribution patterns within the atmosphere. In fact, previous studies have highlighted the importance of clouds in deep atmospheric heating (\citealt{komacekPatchyNightsideClouds2022}). A high cloud opacity on the colder nightside, together with a lower opacity on the dayside due to the lack of clouds and TiO/VO, could potentially lead to an inhomogeneity heating effect rather than a cooling effect. Additionally, our study adopted a simplified drag scheme. For a more accurate understanding of drag effects on internal cooling, future research should consider more realistic approaches, such as the active magnetic drag (e.g., \citealt{rogersMagneticEffectsHot2014,beltzExploringEffectsActive2022}).

The treatment of the lower bottom boundary condition in this study is subtle. There is no solid surface on giant planets, and thus there is no consensus on how to appropriately specify the lower boundary condition in a dynamical model. In this work, we fixed the temperature to mimic a uniform interior entropy in the deep atmosphere and set a sponge layer to dampen the winds at the bottom. In other words, we allowed energy exchange but no mass and momentum transfer between the simulating domain and the underlying interior. Although our model covers a deep convective zone at 200 bars and captures some dynamic heat transport in the weather layer above this level, the overall circulation on a hot Jupiter might potentially reach deeper. Recent studies have found that wind can extend deep on Jupiter (e.g., \citealt{kaspiJupiterAtmosphericJet2018,guillotSuppressionDifferentialRotation2018}). If the primary reason for the end of zonal wind movement on gas giants is the breaking effect of their magnetic fields, then a higher rate of ionization in the atmosphere of a hot Jupiter might point to more shallow circulation, assuming the magnetic field is the same (\citealt{zhangAtmosphericRegimesTrends2020}). In that case, our simulations might have captured the deep wind scenario on hot Jupiters. However, if the circulation on hot Jupiters can extend deeper and carries both mass and energy (like the 2D model setup in \citealt{tremblinAdvectionPotentialTemperature2017}), could the heat flux be transported inward more efficiently? Future simulations should use different methods to investigate how the bottom boundary condition impacts the results.

Planetary atmospheres are inherently 3D. This study demonstrates how the 3D structure of hot Jupiter atmospheres can significantly impact interior heat transport and global-mean quantities in ways that are not captured by 1D models. The inhomogeneity effect suggests the nonlinear dependence of many observable quantities, like internal heat flux, on atmospheric parameters like opacity. While some may consider atmospheres on giant planets as 2D due to their large aspect ratios, this study shows that hot Jupiter atmospheres are not 2D because the convective atmosphere interacts with the radiation-circulation zone above, resulting in a very different pattern of heat flux transport regimes across different zones from the bottom to the top of the atmosphere. Consequently, the internal heat flux at TOA is substantially different from their uniform distribution at the bottom. Thus, it is critical to understand the interior heat transport and inhomogeneity effect in the context of a 3D atmosphere.

This work serves as a bridge between decadal-long 3D atmospheric modeling and traditional 1D interior modeling approaches, highlighting the need for accurate simulations of the atmosphere to appropriately set the outer boundary conditions of planetary evolution. With the powerful observational capabilities of the JWST and ground-based high-resolution spectroscopy, the 3D structure of exoplanetary atmospheres, their surface inhomogeneity, and the horizontal distribution of internal heat flux may be revealed. A 3D weather model of hot Jupiters is a critical tool to completely understand the formation and evolution of these exotic planets.

\section{acknowledgments}

We acknowledge the helpful comments from the reviewer and discussions with Thaddeus Komacek, Daniel Thorngren, and Jo\~{a}o Mendon{\c c}a. X.Z. is supported by NASA Exoplanet Research Grant 80NSSC22K0236 and NASA Interdisciplinary Consortia for Astrobiology Research (ICAR) grant 80NSSC21K0597. We acknowledge computational resources at NASA Pleiades supercomputer and Cluster Lux at UC Santa Cruz. This work was inspired by the last in-person conversation between X.Z. and Dr. Adam P. Showman (1969-2020) about the heat flow on tidally locked exoplanets at the AGU Fall Meeting 2019. We also acknowledge the assistance provided by the OpenAI language model.

\textit{Software}: HARP has been integrated into the SNAP model as a radiative transfer module. The combined SNAP+HARP model, which is open-source, can be accessed freely. The complete version of the source code for this research has been uploaded to Zenodo at DOI:10.5281/zenodo.8176442. The development version of this project, now called CANOE (Comprehensive Atmosphere N'Ocean Engine), is hosted on GitHub at https://github.com/chengcli/canoe.

%

\vspace{5mm}


\appendix
\section{Modified Plane-parallel Radiative Transfer Solver for an extended Atmosphere in Spherical symmetry} \label{app:rte}

Here we compare our modification of the plane-parallel (PP) radiative transfer equation (RTE) with the RTE for a spherical, extended atmosphere to understand the essential assumptions. Physically, there are two main differences along the radial direction between the spherical RTE and the PP RTE. First, in a spherical, extended atmosphere, each atmospheric ``column" is essentially a ``wedge" of a constant solid angle in which the local area changes with altitude. In contrast, in the PP case, the surface area in the column is constant with altitude. Second, the local wedge surface at a radius $r$ on a sphere has a certain curvature. The zenith angle of the light rays changes direction to the surface normal with altitude, and the diffusive flux should account for intensities from other directions. On the other hand, the local surface is assumed to be flat in the PP columns, and the light rays do not change direction during transport, simplifying the calculation.

For simplicity, we assume that the radiation field is axisymmetric, meaning the intensity only depends on the zenith angle $\theta$ and not on the azimuth angle $\phi$. To account for the spherical nature, the radiative transfer equation (RTE) in the spherically symmetric atmosphere is given by \citep{chandrasekharRadiativeEquilibriumExtended1934, chandrasekharRadiativeEqilibriumStellar1945, mihalasFoundationsRadiationHydrodynamics1999}:

\begin{equation}
\mu \frac{dI(r, \mu)}{dr} + \frac{1-\mu^2}{r}\frac{\partial I(r, \mu)}{\partial \mu}= -\kappa\rho \left(I(r, \mu) - S(r, I, \mu)\right), \label{srte}
\end{equation}

Here, $I(r, \mu)$ is the specific intensity of radiation at the radius $r$ from the planetary center in direction $\mu = \cos{\theta}$. $\kappa$ is the extinction coefficient and $\rho$ is the density. The optical thickness $d\tau$ can be defined as $d\tau=-\kappa\rho dr$. $S=(1-\omega)B+\omega S_s $ is the source function, where $B$ is the thermal radiation and $S_s$ is the scattering source, and $\omega$ is the single scattering albedo.

The second term on the left-hand side of \Cref{srte} accounts for the intensity gradient to the zenith angle from the spherical geometry. If we omit that term, the equation is reduced to the plane-parallel RTE (e.g., \citealt{goodyAtmosphericRadiationTheoretical1989, guillotRadiativeEquilibriumIrradiated2010}):

\begin{equation}
\mu \frac{dI(r, \mu)}{dr} = -\kappa\rho (I-S). \label{prte}
\end{equation}

Our proposed modification of the PP RTE is to introduce a transform:
\begin{subequations}
\begin{align}\label{tfeqn}
I&\rightarrow r^2I,\\
B&\rightarrow r^2B. 
\end{align}
\end{subequations}
Then \Cref{prte} becomes:
\begin{equation}
\frac{\mu}{r^2}\frac{d(r^2I)}{d r} = -\kappa\rho(I-B). \label{newprte} 
\end{equation}

This correction mainly accounts for the area change in the radial direction, which is important for energy conservation for extended atmospheres but was neglected in previous studies. However, it has been known that there is no ``plane-parallel" limit of the spherical RTE that can be reduced to the exact form of the PP RTE (e.g., \citealt{hummerRadiativeTransferSpherically1971}). To understand the new PP equation \cref{newprte}, we take the moment method to explore the behavior of the mean intensity $J$ (zeroth moment of the specific intensity $I$), net flux $H$ (first moment), and the second moment $K$ defined as:
\begin{equation}
{J,H,K} = \int_{-1}^{1} {1,\mu,\mu^2} I d\mu.
\end{equation}

Taking the first and second moments of the spherical RTE \cref{srte}, we obtain:
\begin{subequations}
\begin{align}
\frac{\partial H}{\partial r}+\frac{2}{r}H&=-\kappa\rho(J-B), \label{momsrte} \\
\frac{\partial K}{\partial r}+\frac{1}{r}(3K-J)&=-\kappa\rho(1-\omega g_0)H,
\end{align}
\end{subequations}
where $g_0$ is the asymmetry factor of the low-order approximation of the phase function in the scattering source function $S_s$ (see details in \citealt{goodyAtmosphericRadiationTheoretical1989,pierrehumbertPrinciplesPlanetaryClimate2010, hengAnalyticalModelsExoplanetary2014}). To derive the above equations, we also used the conservative nature of the scattering source $S_s$ that disappeared in the first moment equation and the isotropic nature of thermal radiation $B$ that vanished in the second moment equation.

For comparison, the moments of the modified PP RTE \cref{newprte} are:
\begin{subequations}
\begin{align}
\frac{1}{r^2}\frac{\partial}{\partial r}(r^2H) &=-\kappa\rho(J - B),\label{momprte} \\
\frac{1}{r^2}\frac{\partial }{\partial r} (r^2K)&= -\kappa\rho(1-\omega g_0)H.
\end{align}
\end{subequations}
One can see that the first-moment equations in the two frameworks (Equation \ref{momsrte} and Equation \ref{momprte}) are identical. However, the second-moment equations are different unless we set $J=K$. This is essentially the assumption we have made (in the two-stream limit).

In the plane-parallel atmosphere, the ratio of the zeroth moment of the specific intensity, $J$, to the second moment, $K$, is a constant value of 3 for an isotropic or collimated radiation field (e.g., \citealt{guillotRadiativeEquilibriumIrradiated2010}). However, in a spherical atmosphere, the value of $J/K$ is not a constant and varies with optical depth and radius. In the radiative equilibrium situation, $J\sim 3K$ near the bottom boundary, where the spherical effect has not fully emerged, and $J\sim K$ near the outer boundary, i.e., small optical depth, far away from the planetary center (e.g., \citealt{chapmanRadiativeTransferExtended1966,unnoEddingtonApproximationGeneralized1976}).

If one adopts the Eddington approximation, the proposed methods to solve the spherical RTE usually add more equations to find a converged relationship among $J$, $H$, and $K$ (e.g., \citealt{hummerRadiativeTransferSpherically1971, unnoEddingtonApproximationGeneralized1976, simonneauRadiativeTransferAtmospheres1976}). The accurate solution for the spherical, extended atmosphere requires a full 3D RTE solver such as a Monte Carlo RT solver (e.g., \citealt{lucyComputingRadiativeEquilibria1999,lee3DRadiativeTransfer2022}).

Although our assumption of $J=K$ is just one asymptotic behavior of the real spherical RTE solution, it might be the simplest solution to satisfy the flux conservation in a spherical, extended atmosphere in a PP RTE solver like HARP before a computationally efficient spherical RTE solver is developed in the GCM. The net flux ($H$) equation is the same as in the spherical RTE. The derivative of $H$ (scaled with the local surface area) $\frac{1}{r^2}\frac{\partial}{\partial \tau}(r^2H)$ is the atmospheric local heating rate, which is used as a radiative energy source in the energy equation. Because SNAP adopts the finite volume scheme, we used the net flux $H$ directly in the energy equation.

To see the difference between the original PP framework (Equation \ref{prte}) and the modified PP framework (Equation \ref{newprte}), we also derived the moments of the original PP RTE:
\begin{subequations}
\begin{align}
\frac{d H}{\partial r} &=-\kappa\rho(J - B), \\
\frac{d K}{\partial r} &= -\kappa\rho(1-\omega g_0)H.
\end{align}
\end{subequations}
Compared with the modified PP framework (\Cref{momprte}), the original PP framework would yield a different heating and cooling rate. In both frameworks, the heating rates can be calculated as $J-B$. Given the same temperature, $B$ is the same, but the mean intensity $J$ would be different. Thus the thermal cooling rate will be different.

Here we consider a simple example of stellar heating rate. Assuming that the thermal source function $B=0$ in the visible band and no atmospheric scattering of the directly incident flux from the star, we can directly solve for the attenuated fluxes in the two frameworks:
\begin{subequations}
\begin{align}
F(\tau) &= -\mu_0 F_0 e^{-\tau/\mu_0}, ~~~~~~~~~~~\text{original PP} \\
F(\tau) &=-\frac{r_0^2}{r^2}\mu_0F_0 e^{-\tau/\mu_0},\label{visflx} ~~~~~~~\text{modified PP}
\end{align}
\end{subequations}

Here, $\tau$ is the optical depth in the visible band, $\mu_0$ is the cosine of the incoming stellar angle, and $F_0$ is the incoming stellar flux at radius $r_0$ at the TOA. In other words, the modified PP framework directly treats the incident power (flux integrated with area) rather than the flux. The heating rates in the $\tau$ coordinate are:
\begin{subequations}
\begin{align}
        \frac{d F}{d \tau} &=  F_0 e^{-\tau/\mu_0}, ~~~~~~~~~~\text{original PP} \\
        \frac{1}{r^2}\frac{d (r^2F)}{d \tau} &=  \frac{r_0^2}{r^2} F_0 e^{-\tau/\mu_0}. ~~~~~~~\text{modified PP}
\end{align}
\end{subequations}

The modified PP RTE yields a larger stellar heating rate than the original PP RTE because the area is smaller as the flux penetrates deeper. This modification is necessary to satisfy energy (or power) conservation, meaning that if we integrate the heating rate from the bottom to the top of the atmosphere and accurately take into account the area of each atmospheric layer, the total radiative energy is the same as the incoming energy received by the planet.

In other words, directly using the stellar heating rate from the PP framework in current GCMs is biased because the total integrated heating rate in the atmospheric column for an extended atmosphere, if the area factor has been appropriately accounted for, must be smaller than the total incoming stellar flux. It can be easily proved that:
\begin{equation}
        \int_0^{r_0}r^2F_0 e^{-\tau/\mu_0}dr < \int_0^{r_0}r_0^2F_0 e^{-\tau/\mu_0}dr = \mu_0r_0^2F_0.
\end{equation}

Our modified PP framework is a straightforward correction of the PP framework that mainly considers energy conservation. However, we did not account for the detailed treatment of the change of incident and emission angles along the light path. For instance, we did not include the limb treatment when $\mu_0$ approaches or is smaller than 0 (nightside). Although the Chapman function, available in the HARP code, or a path-correction as suggested by \citet{liEffectiveSolarPathlength2006} could be used to correct for this, we did not perform this correction in this study. We wanted to ensure that the global-mean stellar energy attenuation in the 3D global simulation is the same as that in the local simulation. In the local simulation, the mean flux is given by the exponential integral function (Equation \ref{lvflx}). In the future, a real 3D RT model, such as the Monte Carlo RT models, should be used to provide a better parameterization for the extended, spherical atmospheres.

\section{Energy Flux Analysis} \label{app:flx}

Here we derive the vertically integrated and horizontally averaged energy equations for the SNAP model, respectively. We first define the 3D velocity vectors $\mathbf{V} = \{u, v, w\}$ and the horizontal velocity vector $\mathbf{u} = \{u, v\}$. The components $u$, $v$, and $w$ represent velocities in the $x$, $y$, and $z$ directions, respectively, in Cartesian coordinates. In spherical coordinates, $u$, $v$, and $w$ correspond to velocities in the azimuthal (east-west $\phi$), polar (north-south $\theta$), and radial (vertical $r$) directions, respectively.

We can then split the divergence operators ($\nabla$) into horizontal ($\nabla_h$) and vertical ($\nabla_v$) components:
\begin{subequations}
\begin{align}
\nabla \cdot \mathbf{V} &=  \underbrace{\frac{\partial u}{\partial x} + \frac{\partial v}{\partial y}}_{\nabla_h \cdot \mathbf{u}} + \underbrace{\frac{\partial w}{\partial z}}_{\nabla_v \cdot w},  ~~~~~~~~~~~~~~~~~~~~~~~~~~~~~~~~~~~~~~~\text{Cartesian coordinates} \\
\nabla \cdot \mathbf{V} &=  \underbrace{\frac{1}{r\sin{\theta}} \frac{\partial u}{\partial \phi} + \frac{1}{r\sin{\theta}} \frac{\partial (\sin{\theta} v)}{\partial \theta}}_{\nabla_h \cdot \mathbf{u}} + \underbrace{\frac{1}{r^2} \frac{\partial (r^2 w)}{\partial r}}_{\nabla_v \cdot w}.  ~~~~~~~\text{Spherical coordinates}
\end{align}
\end{subequations}

The energy equation in the non-hydrostatic (fully compressible) atmosphere is then represented as (e.g., see Equation 1c in \citealt{zhangAtmosphericRegimesTrends2020}):
\begin{equation}
\frac{\partial (\rho e + \rho \Phi)}{\partial t} + \nabla_h \cdot [(\rho e + p + \rho \Phi) \mathbf{u}] + \nabla_v \cdot [(\rho e + p + \rho \Phi)w] + \nabla_v \cdot F_{rad} = 0,
\label{vector-energy-eq}
\end{equation}
where $t$ is time and $p$ is pressure. $\Phi$ is the geopotential defined as $g=-\nabla_v \Phi$, where $g$ is gravity in the radial direction. In this equation, we neglect thermal conduction. The net radiative flux in the radial direction is $F_{rad}=F_{\rm v}+F_{\rm IR}$, where $F_{\rm v}$ and $F_{\rm IR}$ are the net fluxes in the visible and infrared bands, respectively. As we neglect scattering in the visible band, $F_{\rm v}$ only has a downward (negative) component. There is a positive sign with $\nabla_v \cdot F_{rad}$ because we define upward flux as positive, and the atmosphere cools down if the upward flux increases with height.

The total energy in the atmosphere can be expressed as $\rho e = \rho c_v T + \rho E_k$, where $\rho c_v T$ is the internal energy and $\rho E_k$ is the kinetic energy, with $E_k = \frac{1}{2}(u^2 + v^2 +w^2)$ being the sum of the squares of the velocity components. Using the ideal gas law $p=\rho R T$ with $R$ being the gas constant and $c_p=c_v+R$ where $c_p$ is the specific heat at constant pressure, we can write $\rho e + p = \rho c_p T + \rho E_k$, which is composed of the sensible heat $\rho c_p T$ and the kinetic energy $\rho E_k$.

In a steady state, the time derivative on the left-hand side of \Cref{vector-energy-eq} vanishes. We also assume zero vertical velocity at the top and bottom boundaries. Taking the vertical integration from the bottom to the top of the atmospheric column, we obtain the vertically integrated energy equation:
\begin{equation}
\langle\nabla_h \cdot \rho(c_p T + \Phi + E_k) \mathbf{u}\rangle + F_{\rm v,\rm TOA} + F_{\rm IR,\rm TOA}-\overline{F_{\rm int}}=0,
\label{vert-int-eq}
\end{equation}
where $\langle\cdot\rangle$ denotes the vertical integration. The energy budget has two components: the dry static energy term $c_p T + \Phi$ and the kinetic energy $E_k$. $F_{\rm v,\rm TOA}$ and $F_{\rm IR,\rm TOA}$ are the local incoming stellar flux and local outgoing IR radiation at the top of the atmosphere, respectively. The integration constant is equal to the domain-averaged internal (IR) heat flux $\overline{F_{\rm int}}$. Although $F_{\rm int}$ is defined as the spatially varying local internal heat flux at the TOA (Equation \ref{finteq}), the domain-averaged value is a constant throughout the atmosphere, as shown in the horizontally averaged energy equation below.

To obtain the horizontally averaged energy flux equation, we first separate the geopotential term from the divergences in \Cref{vector-energy-eq}. Because $g=-\nabla_v \Phi$ only exists in the radial direction, we have:
\begin{equation}
\nabla_h \cdot (\rho \Phi \mathbf{u}) + \nabla_v \cdot (\rho \Phi w) = \Phi\nabla\cdot (\rho \mathbf{V}) + \rho w\cdot\nabla\Phi=\rho gw,
\end{equation}
where $\nabla\cdot (\rho \mathbf{V})=0$ due to the mass continuity. The energy equation \Cref{vector-energy-eq} can be written in a form with gravity as an external forcing (e.g., \citealt{liSimulatingNonhydrostaticAtmospheres2019,geGlobalNonHydrostaticAtmospheric2020}):
\begin{equation}
\frac{\partial (\rho e)}{\partial t} + \nabla_h \cdot [(\rho e + p) \mathbf{u}] + \nabla_v \cdot [(\rho e + p)w] +\rho g w + \nabla_v \cdot F_{rad}=0. \label{eneeqn}
\end{equation}

In the steady state, several terms in the above equation vanish in the domain and temporal average, including the time derivative, horizontal divergence, and the potential energy term $\rho g w$. Using $\rho e = \rho c_v T + \rho E_k$ and $F_{rad}=F_{\rm v}+F_{\rm IR}$, we can obtain the horizontally averaged vertical flux divergence equation:
\begin{equation}
\nabla_v \cdot \left(\overline{\rho c_p wT} + \overline{\rho w E_k} + \overline{F_{\rm v}}+\overline{F_{\rm IR}}\right)=0,
\label{hori-avg-eq}
\end{equation}

where $\overline{A}$ denotes the horizontally averaged quantity of $A$ over the horizontal domain and time. We can then vertically integrate \Cref{hori-avg-eq} to obtain the global-mean flux balance equation. Because the divergence operators are different between the local and global models, the integration is different. In the local model, the integration constant is the mean internal heat flux $\overline{F_{\rm int}}$ that is conserved in the atmosphere. But in an extended, spherical atmosphere where the surface area changes with height, the mean internal luminosity (or emitting power, in units of watts), rather than the internal flux (luminosity per area), is conserved.

To facilitate the comparison between the local and global models, we normalized the fluxes in \Cref{hori-avg-eq} by the globally averaged incoming stellar flux $\sigma T_{eq}^4$ in the local model and by $r_0^2 \sigma T_{eq}^4/r^2$ (scaling the energy power) in the global model. Then we integrate the equation with height in the local model and over the spherical shell in the global model. The normalized mean energy flux equations in the two models now have the same formalism:
\begin{equation}
\overline{\rho c_p wT} + \overline{\rho wE_k} + \overline{F_{\rm v}}+\overline{F_{\rm IR}}=\overline{F_{\rm int}}.
\label{flx-eq}
\end{equation}

We have used the same variables in \Cref{hori-avg-eq} for the normalized fluxes for convenience. To avoid confusion, we \textit{only} used the normalized fluxes for the domain-averaged fluxes in this study. The normalized internal heat flux $\overline{F_{\rm int}}$ is a constant in both models and is determined by the boundary condition. Because the vertical velocities are zero at the top and bottom boundaries and the normalized stellar flux $\overline{F_{\rm v}}=-1$ (downward) at the upper boundary, we have
\begin{equation}
\overline{F_{\rm int}}=\overline{F_{\rm IR,\rm TOA}}-1.
\label{toaint-eq}
\end{equation}

If we take the horizontal average of \Cref{vert-int-eq}, the horizontal divergence vanishes, and we can also reproduce \Cref{toaint-eq}. If the hot Jupiter is in energy equilibrium, the unknown heating heat source must compensate for the outgoing internal heat flux. Thus the required heating efficiency $\eta$ is equal to the normalized internal heat flux $\overline{F_{\rm int}}$ (Equation \ref{eff}).



\end{document}